\documentclass[useAMS,usenatbib]{mn2e}
\usepackage{subfigure}
\usepackage{captcont}
\usepackage{longtable}
\usepackage{graphicx}
\usepackage{natbib}

\title[Survey of O~{\small VI} in the LMC]{Survey of O~{\small VI} absorption in the Large Magellanic Cloud}
\author[A. Pathak et al.]{A. Pathak\thanks{e-mail: amitp@iiap.res.in; amitpah@gmail.com}, A. C. Pradhan, N. V. Sujatha and J. Murthy\\
Indian Institute of Astrophysics, Koramangala II Block, Bangalore 560034, India}

\begin{document}

\date{}

\pagerange{}

\maketitle
\label{firstpage}

\begin{abstract}
We present a survey of interstellar O~{\small VI} absorption in the Large Magellanic Cloud (LMC) towards 70 lines of sight based on {\it Far Ultraviolet Spectroscopic Explorer (FUSE)} observations. The survey covers O~{\small VI} absorption in a large number of objects in different environmental conditions of the LMC. Overall, a high abundance of O~{\small VI} is  present in active and inactive regions of the LMC with mean log~N(O~{\small VI}) = 14.23 atoms cm$^{-2}$. There is no correlation observed between O~{\small VI} absorption and emissions from the hot gas (X-ray surface brightness) or the warm gas (H$_{\alpha}$ surface brightness). O~{\small VI} absorption in the LMC is patchy and the properties are similar to that of the Milky Way (MW). In comparison to the Small Magellanic Cloud (SMC), O~{\small VI} is lower in abundance even though SMC has a lower metallicity compared to the LMC and the MW. We present observations in 10 superbubbles of the LMC of which we detect O~{\small VI} absorption in 5 superbubbles for the first time and the superbubbles show an excess O~{\small VI} absorption of about 40\% compared to non-superbubble lines of sight. We have also studied the properties of O~{\small VI} absorption in the 30 Doradus region. Even though O~{\small VI} does not show any correlation with X-ray emission for the LMC, a good correlation between log~N(O~{\small VI}) and X-ray surface brightness for 30 Doradus region is present. We also find that O~{\small VI} abundance decreases with increasing distance from the star cluster R136.
\end{abstract}


\section{Introduction}

The interstellar medium (ISM) of the Milky Way (MW) and other galaxies is a complex mix of gas and dust. The processes involved in maintaining the mass, energy, and ionization balance of the ISM are not properly understood. High resolution ultraviolet ({\it UV}) spectra provides information about these processes as many absorption lines of atoms, ions and molecules are present in the {\it UV} band of the electromagnetic spectrum, one of the most important of which is O$^{+5}$ (O~{\small VI}), a diagnostic of temperatures of about $3 \times 10^{5} $~K \citep{Cox05}. Such temperatures are found at the interface of hot (T $> 10^{6}$ K) and warm (T $\sim 10^{4}$ K) ionized gas in the ISM. Thus, O~{\small VI} absorption lines at 1031.9 \AA\ and 1037.6 \AA\ are crucial diagnostics of the energetic processes of interface environments in the ISM of galaxies. The gas at such temperatures is cooling radiatively and the cooling is essentially independent of density, metallicity and the heating mechanism \citep{Edgar86, Heckman02}. O~{\small VI} formation by photo-ionization is unable to explain the observed abundances, given the energy of photons needed to get such high ionization (114 eV). O~{\small VI} is mostly produced by shock heating and is collisionally ionized \citep{Indebetouw04}.

Previous studies of O~{\small VI} have been limited to observations by the {\it Copernicus} satellite \citep{Jenkins78a, Jenkins78b} and the {\it Hopkins Ultraviolet Telescope} \citep{Dixon96}. \citet{Shelton94} concluded that the hot gas exists in discrete regions rather than being continuously present in the ISM. The launch of {\it Far Ultraviolet Spectroscopic Explorer (FUSE)}; \citep{Moos00, Sahnow00} enabled a wider and more descriptive study of O~{\small VI} absorption and emission in the ISM and the intergalactic medium (IGM). With a spectral resolution of ~ 20,000, {\it FUSE} has been able to resolve fine details of O~{\small VI} in many different environments. {\it FUSE} observed O~{\small VI} absorption lines in the local ISM of the MW \citep{Savage00, Wakker03, Oegerle05, Savage06, Welsh08}, disk of the MW \citep{Bowen08}, halo of the MW \citep{Savage03}, the Large Magellanic Cloud (LMC; \citet{Howk02a, Lehner07}), the Small Magellanic Cloud (SMC; \citet{Hoopes02}), starburst galaxies \citep{Grimes09}, IGM \citep{Danforth05, Danforth06}, etc. Apart from absorption studies, {\it FUSE} has also recorded O~{\small VI} spectra in emission from observations of diffuse ISM in the MW \citep{Shelton01, Shelton02, Dixon06, Dixon08} and superbubbles in the LMC \citep{Sankrit07}. These studies have not only augmented our knowledge about the formation and distribution of O~{\small VI} in the MW and the Magellanic Clouds but have also helped in better understanding of the complexities of the ISM.

Owing to the contiguity to the MW ($\sim$ 50 kpc; \citet{Feast99}) and being nearly a face-on galaxy with a low inclination angle ($\sim$ 35$^{\circ}$; \citet{Marel01}), the LMC has been the subject of numerous studies to understand and interpret the properties of ISM. Recent studies of the ISM in the LMC include observations of diffuse UV emission \citep{Cole99a, Cole99b, Pradhan10}, the {\it SPITZER} infrared dust survey (SAGE; \citet{Meixner06, Bernard08}), the HI survey \citep{Kim03}, the H$_{\alpha}$ survey \citep{Gaustad01}, the O~{\small VI} distribution \citep{Howk02a}, and the survey of hot gas in the X-ray bands \citep{Snowden94}. More recently, \citet{Lehner07} analyzed the absorption of O~{\small VI}, C {\small IV}, Si {\small IV} and N {\small V} ions towards four early type stars in the LMC and provide crucial details about the environments that may be probed through the study of these ions.

\citet{Howk02a} surveyed the distribution and kinematics of O~{\small VI} towards 12 early type stars in the LMC, which was very selective and the targets were restricted to Wolf-Rayet stars and O-type stars of spectral types O7 and earlier. Here we report an extensive survey of O~{\small VI} absorption at 1032 \AA\ in the LMC. The observation log is given in Table \ref{Tab1}. The results presented in this survey become important as they not only provide a detailed mapping of the interface gas traced by O~{\small VI} but also throw light on the small scale structure of the O~{\small VI} distribution in various regions of the LMC. Fig. \ref{Fig1} shows the lines of sight and different regions of the LMC including the superbubbles (marked by circles) that have been studied in this survey.

\section{Observations and data analysis}

\subsection{{\it FUSE} data analysis and possible contamination}

The {\it FUSE} mission and its operations are described by \citet{Moos00} and \citet{Sahnow00}. Observations are made through one of 3 apertures: the HIRS (\(1.25''\) $\times$ \(20''\)); the MDRS (\(4''\) $\times$ \(20''\)); and the LWRS (\(30''\) $\times$ \(30''\)), with all three obtaining data simultaneously. Depending on the coating of the spectrograph, observations are possible through SiC and LiF channels that are further divided into eight different segments; the SiC 1A, SiC 2A, SiC 1B and SiC 2B segments covering the wavelength range 905 to 1105 \AA\ and LiF 1A, LiF 2A, LiF 1B and LiF 2B segments covering the wavelength range 1000 to 1187 \AA. The data from SiC 2B segment have been known to suffer from a fixed noise. The sensitivity of LiF 1A segment near 1032 \AA\ is almost double that of other segments and therefore, we have used only the LiF 1A observations \citep{Sahnow00}. Most of the data are from the large aperture LWRS with a few from the MDRS aperture.

The fully calibrated {\it FUSE} spectra were downloaded from the Multimission Archive at STScI (MAST) processed by the latest {\it FUSE} data reduction pipeline (CALFUSE version 3.2; \citet{Dixon07}). {\it FUSE} made more than 600 pointings in and around the LMC. Several of the observations were rejected initially by just looking at the spectra (with non-existent or low signal-to-noise of O~{\small VI} absorption). 70 unique {\it FUSE} targets were selected based mainly on the simplicity of the continuum fitting in the vicinity of O~{\small VI} (at 1031.9 \AA). O~{\small VI} absorption for 1 of the sightline has been reported by \citet{Friedman00}, 11 have been covered by \citet{Howk02a} survey, 3 by \citet{Danforth06a} and 1 by \citet{Lehner07}. Spectral types and other stellar information was taken from \citet{Danforth02} and \citet{Blair09}. We have downgraded all the spectra reported here to 35 km~s$^{-1}$ to have a higher signal-to-noise. This has been done for all the spectra irrespective of the quality to maintain uniformity in the data analysis procedure. 

Fitting the stellar continuum in the neighbourhood of the O~{\small VI} absorption profile has been discussed in detail by \citet{Friedman00} and \citet{Howk02a}. \citet{Howk02a} have limited their study to early type stars based on the fact that these stars have completely developed O~{\small VI} P Cygni profiles that are easier to fit as the early type stars have a high mass loss rate and have minimal wind variations. \citet{Lehner01,Lehner03} have shown that for Galactic sight lines, the stellar wind variability has negligible effect on O~{\small VI} absorption but may introduce substantial errors towards targets in the Magellanic clouds. \citet{Lehner01} did find variation in the equivalent width and column density towards a LMC star when estimated at two different times. This warrants for extra care while fitting the continuum for Magellanic cloud targets.

As discussed above, for the LMC, the estimation of stellar continuum in the vicinity of O~{\small VI} absorption is not trivial. Our background targets are mostly early O and B type stars with several Wolf-Rayets. These have been selected based on the fact that the the continuum near the O~{\small VI} absorption at 1031.9 \AA\ is simple to fit. A few of the lines of sight do show a complex behaviour near O~{\small VI} absorption (for e.g., Sk-67D05, Sk-67D168, Sk-70D115, etc.). For such exceptional lines of sight, the complexity in the continuum fitting is due to a local dip or a sudden rise near the O~{\small VI} absorption and these targets needed a comparatively higher order polynomial for the fitting. For a flavour of the complexity involved, sample continuum fitting for three lines of sight, BI13, Sk-70D115 and Sk-67D168 are shown in Fig. \ref{Fig2}. Following the fitting procedure of \citet{Howk02a} and \citet{Sembach92}, the local stellar continuum was estimated for all the targets and were fitted by a Legendre polynomial fit of low order ($\leq 5$). Several continua were tested and the uncertainties involved for the complete data set were used in the measurement of the O~{\small VI} column densities. The normalized spectra in the vicinity of O~{\small VI} absorption are presented in Fig. \ref{Fig3}.

The O~{\small VI} absorption may be contaminated by absorption from molecular hydrogen, however, it is minimal in the LMC velocity range. This is because the molecular fraction of H$_2$ in the LMC is only about 12\% of the molecular fraction in the Galactic disk \citep{Tumlinson02}. The closest absorption line of H$_2$ to O~{\small VI} absorption in the LMC lies at 1032.35 \AA\ (at v$_{lsr} \sim$ +123 km~s$^{-1}$), which is due to (6--0) R(4) transition. An estimation of possible contamination by H$_2$ towards 12 lines of sight in the LMC has been done by \citet{Howk02a} where they find that H$_2$ absorption does not affect the LMC O~{\small VI} column densities significantly. The H$_2$ absorption feature are usually not a problem for measurements of O~{\small VI} absorption at velocities more than 20 km~s$^{-1}$ apart (\citet{Savage03} and references therein). We estimate the contamination from H$_2$ lines by fitting the (6--0) P(3) H$_2$ transition for the MW and (6--0) R(4) H$_2$ transition for the LMC (Fig. \ref{Fig4}). From this, we are able to estimate the contamination due to (6--0) R(4) MW H$_2$ line and (6--0) P(3) LMC H$_2$ line. Noticeable contamination is seen for the sightlines Sk-71D45 and Sk-67D250 which is well within the error bar of the derived column densities. An example of a sightline is also shown in Fig. \ref{Fig4} in which the LMC O~{\small VI} absorption is free from any contamination from molecular hydrogen. Owing to the negligible contribution, we have not excluded the contamination by H$_2$ in the O~{\small VI} measurements for the LMC.

 Another possible contaminant is the (6--0) R(0) transition of the HD molecule at 1031.91 \AA\ that overlaps with the MW O~{\small VI} absorption. Several lines of HD molecule is covered by {\it FUSE} (see \citet{Sembach99} for a complete list). H$_2$ column densities are significantly low in the LMC and contamination due to molecular hydrogen lines is non-significant for O~{\small VI} measurements in the LMC. Absorption due to HD is about 4 orders weaker compared to H$_2$ transitions. Therefore, we neglect any contribution from HD molecule to the column densities of O~{\small VI} in the LMC.

\subsection{Measurement of O~{\small VI} column densities}

The measurement of equivalent widths and column densities of O~{\small VI} absorption for all the lines of sight were done following \citet{Savage91, Sembach92} and \citet{Howk02a}. This apparent optical depth technique \citep{Savage91} is now commonly used in the analysis of interstellar absorption lines and is applicable to cases with non-saturated absorption profiles. Briefly the technique uses an apparent optical depth in terms of velocity, i.e., an instrumentally blurred version of the true optical depth, given as

\begin{equation}
\tau_a(v) = ln[I_{o}(v)/I_{obs}(v)],
\end{equation}
where $I_o$ is the estimated continuum intensity and $I_{obs}$ is the intensity of the absorption line as a function of velocity. If the resolution of the instrument is very high compared to the FWHM of the absorption line, the apparent optical depth is a very good representation of the true optical depth. The apparent column density ($N_a(v)$ [atoms cm$^{-2}$ (km~s$^{-1}$)$^{-1}$]) is calculated by the following relation

\begin{equation}
N_a(v) = \frac{m_e c \tau_a(v)}{\pi e^2 f \lambda} = 3.768 \times 10^{14} \frac{\tau_a(v)}{f \lambda},
\end{equation}
where $\lambda$ is the wavelength (in \AA) and {\it f} is the oscillator strength of the atomic species (for O~{\small VI}, {\it f} value of 0.1325 has been adopted from \citet{Yan98}). Similar to \citet{Howk02a}, we find that the 1032 \AA\ O~{\small VI} profile is broad and is fully resolved by {\it FUSE}. However, the weaker absorption of the O~{\small VI} doublet at 1037.6 \AA\ is found to be inseparable from the CII* and H$_2$ absorption.

The details of the apparent optical depth measurements are listed in Table \ref{Tab2}. The overlap of the O~{\small VI} absorption of the MW and the LMC does not allow a precise measurement of column densities for the LMC. For the lines of sight (Sk-65D21, Sk-67D69, Sk-67D105, HV2543, Sk-66D100, HV5936, Sk-67D211, Sk-69D220, Sk-66D172, Sk-68D137, D301-1005, and D301-NW8) where the LMC O~{\small VI} is distinct from the MW O~{\small VI}, the limits of velocities over which integration was performed were easy to obtain. Taking cue from these lines of sight, we estimate the lower velocity limit for the lines of sight where the LMC O~{\small VI} was not separated from the MW O~{\small VI}. The errors in the equivalent widths and column densities are 1$\sigma$ uncertainties derived using the uncertainty in the {\it FUSE} data and the fitting procedures. The principal contributor to these errors is the ambiguity in the continuum fitting of several of the spectra.

\section{Distribution and properties of O~{\small VI} in the LMC}

\subsection{Abundance and  linewidth of O~{\small VI}}

The LMC spectra were selected based on the quality of the spectra around O~{\small VI} spectral feature. The survey presented here has more-or-less complete coverage of the well known regions of the LMC such as 30 Doradus, N11, LMC 4, etc., and thus, is very useful in studying the variation of O~{\small VI} at small scales as well. We find a significant amount of variation in the O~{\small VI} column densities in a single region and in different regions of the LMC.

Our data cover a wide range of targets; O type stars, B type stars, Wolf-Rayet objects, supernova remnants, etc. The complete list of the background targets is provided in Table \ref{Tab1}. These observations give fine details of the small scale structure of O~{\small VI} column density in the LMC probing to a scale of $\sim$ 10 pc in some regions. We find the O~{\small VI} absorption in the LMC to be very patchy as is true for the MW. The O~{\small VI} column density varies from the lowest value of 5.28 $\times 10^{13}$ atoms cm$^{-2}$ near a HII region (DEM 7) to the highest value of 3.74 $\times 10^{14}$ atoms cm$^{-2}$ just north of the LMC bar. \citet{Howk02a} report the distribution of O~{\small VI} in the LMC to be patchy as well with the log~N(O~{\small VI}) value in the LMC varying from 13.9 to 14.6 atoms cm$^{-2}$ and a mean of 14.37 atoms cm$^{-2}$. Their data did not provide any small scale variation in the column density of O~{\small VI} with the smallest scale probed is about 450--500 pc.

Given the coverage of our data, we have studied the variation in O~{\small VI} absorption in different regions of the LMC. We find that the 30 Doradus and SN 1987 A regions dominate in terms of the abundance of O~{\small VI}. The mean and median values of log~N(O~{\small VI}) for targets in and around 30 Doradus and SNR 1987A are 14.30 and 14.27 atoms cm$^{-2}$ respectively. 30 Doradus is the largest HII region of LMC with a dense concentration of early type massive stars and has thus, attracted extensive research in different wavelength bands \citep{Walborn92, Parker93, Malumuth94, Walborn97, Townsley06, Indebetouw09}. We discuss in detail about the O~{\small VI} distribution and its properties in 30 Doradus region in section 5.

N11 region, which is associated with several OB associations and has a superbubble at its center, shows a high value of O~{\small VI} column density. The mean of log~N(O~{\small VI}) in the N11 region is 14.21 atoms cm$^{-2}$.

Another interesting region is the LMC 4 Supergiant shell that includes the Shapley Constellation III, which is one of the largest region associated with star formation \citep{Dopita85, Dolphin98}. The log~N(O~{\small VI}) value in LMC 4 supershell varies from a minimum of 13.86 atoms cm$^{-2}$ to a maximum of 14.45 atoms cm$^{-2}$ with a mean value of 14.20 atoms cm$^{-2}$ and a median value of 14.25 atoms cm$^{-2}$. Other regions of the LMC also show patchiness in the O~{\small VI} distribution. Statistically, the mean of the O~{\small VI} column density in the LMC is 1.88 $\times 10^{14}$ atoms cm$^{-2}$ which is lower than 2.34 $\times 10^{14}$ atoms cm$^{-2}$ given by Howk et al. (2002) for 12 lines of sight. The difference is most likely due to the wide coverage of background targets in our data. The median value of the O~{\small VI} column density is 1.66 $\times 10^{14}$ atoms cm$^{-2}$ which also is less than the value reported by \citet{Howk02a}. Overall we find that there is an ubiquitous presence of O~{\small VI} throughout the LMC.

The kinematics of O~{\small VI} in the LMC is difficult to study because of the ambiguity in separating the LMC O~{\small VI} absorption from the MW absorption. For some of the {\it FUSE} targets, i.e., for Sk-65D21, Sk-67D69, Sk-67D105, HV2543, Sk-66D100, HV5936, Sk-67D211, Sk-69D220, Sk-66D172, Sk-68D137, D301-1005, and D301-NW8, O~{\small VI} absorption at LMC velocities are distinct from the MW (Fig. \ref{Fig3}). Linewidth for the LMC O~{\small VI} profiles are obtained with relatively less error for these lines of sight. The LMC O~{\small VI} absorption profiles have all been fitted with a single Gaussian. The corresponding FWHM for these lines of sight are 55, 85, 78, 90, 90, 113, 100, 85, 105, 94, 107, and 91 km~s$^{-1}$ respectively. The temperature range estimated from these widths is T~$\sim$ 1 $\times$ 10$^6$ -- 5 $\times$ 10$^6$ K. O~{\small VI} abundance is maximum at a temperature of 3 $\times$ 10$^5$ K, thus, higher FWHM of these profiles is probably due to other broadening mechanisms such as more than one velocity component and/or collision and turbulence. This also represents the kinematic flow structure of O~{\small VI} in the LMC.

The linewidth for the MW O~{\small VI} profiles are narrower than the LMC profiles \citep{Savage03, Oegerle05, Savage06}. For the Galactic halo, \citet{Savage03} report a range for $\sigma$ (linewidth) from 16 to 65 km~s$^{-1}$ (corresponding FWHM range is 38--153 km~s$^{-1}$). \citet{Oegerle05}, for the local ISM, report average $\sigma$ to be 16 km~s$^{-1}$ (FWHM = 38 km~s$^{-1}$) while \citet{Savage06} report $\sigma$ values ranging from 15 km~s$^{-1}$ to 36 km~s$^{-1}$ (FWHM ranging from 35 km~s$^{-1}$ to 85 km~s$^{-1}$). The linewidth of the SMC O~{\small VI} absorption profile are comparable to that of the LMC. The FWHM range for the SMC O~{\small VI} absorption is from 82 to 115 km~s$^{-1}$ with a mean of 94 km~s$^{-1}$ \citep{Hoopes02}. We obtained a mean FWHM value of 91 km~s$^{-1}$ for the LMC selected lines of sight. \citet{Howk02a} have compared O~{\small VI} absorption profiles of the LMC with Fe II absorption and find that the Fe II profiles are much narrower suggesting that the thermal broadening effect for O~{\small VI} absorption is much more significant.

\subsection{Comparison with the MW and the SMC}

The MW and the SMC offer a different ISM environment compared to the LMC especially due to the difference in the metallicity. The absorption profiles of the MW O~{\small VI} are different from the LMC O~{\small VI} and sometimes these profiles are difficult to separate in an unambiguous manner, thus, a comparison of the kinematics is not possible. \citet{Howk02a} have compared the O~{\small VI} absorption with Fe II absorption at 1125.45 \AA\ (fig. 9 in \citet{Howk02a}). While, the Fe II profiles for the MW and the LMC are clearly separated from each other, the O~{\small VI} absorption suffers an overlap. This is due to the difference in width of the two absorption. The overlap of the MW O~{\small VI} absorption profile with the LMC O~{\small VI} absorption constrained \citet{Howk02a} to arrive at a definite conclusion about the existence of outflows from the LMC. Following the discussion about the distribution of O~{\small VI} in the LMC in the previous section, we now present an overview of O~{\small VI} in the MW and the SMC.

O~{\small VI} in the MW has been extensively studied since the launch of {\it FUSE} \citep{Savage00, Howk02b, Wakker03, Savage03, Oegerle05, Savage06, Bowen08}. \citet{Savage00} were the first to study the O~{\small VI} absorption in the disk and halo of the MW as seen towards 11 extragalactic objects (active galactic nuclei) using {\it FUSE}, confirming the large scale presence of hot gas in the halo \citep{Spitzer56}. The authors find that log N$_{\perp}$(O~{\small VI}) varies from 13.80 to 14.64 atoms cm$^{-2}$ and the distribution of O~{\small VI} is quite patchy. To compare with the projected O~{\small VI} column density on the plane of the MW, we calculated the projection of O~{\small VI} column on to the plane of the LMC. Taking the inclination angle of the LMC to be 33$^{\circ}$, we find the mean value of log N$_{\perp}$(O~{\small VI}) $\equiv$ log~N(O~{\small VI}) cos~$\theta$ = 14.16 atoms cm$^{-2}$. \citet{Savage00} quote a mean value of 14.29 atoms cm$^{-2}$ for their sample. The median value of our sample is 14.14 atoms cm$^{-2}$, while for the \citet{Savage00} sample, it is 14.21 atoms cm$^{-2}$.

\citet{Savage03} report {\it FUSE} observations of O~{\small VI} absorption towards 100 extragalactic lines of sight to study the properties and distribution of O~{\small VI} in the galactic halo. The average log N (O~{\small VI}) for the complete sample is 14.36 atoms cm$^{-2}$ while log N (O~{\small VI}) sin~$|b|$ value for the complete sample is 14.21 atoms cm$^{-2}$. The results reveal that there are substantial differences in the values of log~N(O~{\small VI}) and log~N(O~{\small VI}) sin$|b|$ in the northern Galactic hemisphere compared to the southern Galactic hemisphere. The patchiness in the distribution of O~{\small VI} absorption is found to be similar over angular scales extending from $\leq$ 1$^{\circ}$ to 180$^{\circ}$. An extensive survey of O~{\small VI} in the MW disk has been reported by \citet{Bowen08} in which the authors have studied O~{\small VI} column density towards 148 early type stars. The correlation between O~{\small VI} column density and effective distance to a star establishes the fact that the O~{\small VI} is interstellar in nature and points to the universal presence of the interstellar phenomena that gives rise to O~{\small VI} throughout the Galaxy.

\citet{Hoopes02} have surveyed O~{\small VI} absorption towards 18 early type stars in the SMC and report a widespread presence of O~{\small VI}. The mean value of log~N(O~{\small VI}) in the SMC is 14.53, which is higher than the LMC and the MW values. The O~{\small VI} column density in the SMC correlates with the distance from NGC 346, a star forming region that shows the highest abundance of O~{\small VI} in the SMC.

\subsection{Comparison with X-ray and H$\alpha$}

The LMC has been the focus of H$_{\alpha}$ and X-ray surveys to search for ionized structures in the ISM. H$_{\alpha}$ surveys have revealed the presence of HII regions, supernova remnants, and large scale structures including superbubbles and super-shells \citep{Davies76}, whereas, X-ray observations have been done to study bright X-ray sources \citep{Trumper91}, the hot gas in the ISM \citep{Wang89, Wang91} and diffuse X-ray emission \citep{Bomans94}. Since O~{\small VI} traces the ISM gas with temperatures $\sim$ 10$^5$K, which is at the interface between hot gas (temperature $\geq$ 10$^6$K) traced by X-rays and warm gas (temperature $\sim$ 10$^4$K) traced by H$_{\alpha}$, correlation between O~{\small VI} abundance and X-ray and H$_{\alpha}$ emissions is expected.

The O~{\small VI} observations cover specific regions of the LMC with varying environmental conditions. To get an idea about the variation of O~{\small VI} column densities with different environments in the LMC, we have overlaid O~{\small VI} column densities as circles on H$_{\alpha}$ image (\citet{Gaustad01}; Fig. \ref{Fig5}). The diameter of the circle is linearly proportional to log~N(O~{\small VI}). Interestingly, it is noted that O~{\small VI} abundance is high in regions with low H$_{\alpha}$ and X-ray emissions, i.e., regions that are relatively inactive. However, regions like superbubbles are O~{\small VI} rich, for e.g., 30 Doradus C and N11. The gross picture suggests that O~{\small VI} does not correlate with either H$_{\alpha}$ or X-ray emissions. We have plotted log~N(O~{\small VI}) against log relative H$_{\alpha}$ and X-ray surface brightnesses to have a better insight. The X-ray surface brightness is obtained from {\it ROSAT} PSPC mosaic image of the LMC covering the energy range 0.5--2.0 keV \citep{Snowden94}. Figures \ref{Fig6} and \ref{Fig7} show the correlation of log~N(O~{\small VI}) with H$_{\alpha}$ and X-ray surface brightnesses for the LMC barring five lines of sight in the X-ray correlation plot. Four of the excluded lines of sight (Sk-67D250, D301-1005, D301-NW8 and Sk-65D63) are not covered by the X-ray observations and one (Sk-69D257) has extremely high X-ray emission (about 2 orders of magnitude higher than other lines of sight). The H$_{\alpha}$ and X-ray surface brightness are measured by re-binning the corresponding images to increase the signal to noise. The lack of correlation is evident in both plots, as also found by \citet{Howk02a}. Since, H$_{\alpha}$ traces warm ISM and star formation, there seems to be no direct relation between O~{\small VI} formation and these processes. A better correlation with X-ray is expected as the hot gas traced by X-ray is presumed to cool through temperatures where O~{\small VI} formation takes place. There is a high abundance of O~{\small VI} in supernova remnants and superbubbles of the LMC, whereas, bright X-ray emission is observed mostly from the supernova remnants of the LMC. X-ray emission associated with supernova remnants in the LMC is a factor of 2 to 3 times greater than the X-ray emission associated with supergiant shells and the Bar in the LMC \citep{Points01}.

\section{O~{\small VI} in superbubbles of the LMC}

Superbubbles may be formed due to the strong stellar wind from young stars and/or supernova explosions forming a local cavity in the surrounding ISM. These are excellent examples of the interaction of young hot stars and the ISM. Superbubbles contain hot gas ($\sim 10^{6} K$) that is heated by shocks created by stellar winds \citep{Castor75, Weaver77, Chu90}, which is an ideal condition for the formation of O~{\small VI}. The LMC hosts more than 20 superbubbles and recently O~{\small VI} has been detected in a superbubble N70 \citep{Danforth06a}, where they have reported around 60\% excess in abundance of O~{\small VI} in comparison to non-superbubble lines of sight in the proximity of N70. The authors conclude that superbubbles act as local O~{\small VI} reservoirs and have a different absorption profile compared to the non-superbubble O~{\small VI} absorption profiles. They find that the O~{\small VI} is formed by the thermal conduction between the interior hot X-ray producing gas and the cool photo-ionized shell of N70. \citet{Sankrit07} have detected O~{\small VI} in emission in several superbubbles in the LMC further emphasizing that superbubbles are important contributors to the overall O~{\small VI} budget.

We have 22 O~{\small VI} observations covering 10 superbubbles. Of the 22 observations, 3 are in N70 (Sk-67D250, D301-1005 and D301-NW8) which have already been reported by \citet{Danforth06a} and 4 are in N144, N204, N206 and N154 respectively which have been discussed by \citet{Howk02a}. We report 15 new observations of O VI absorption in the superbubbles 30 Doradus C, N158, N11, N51, and N57. Significant variation in O~{\small VI} abundance exists towards these lines of sight. The minimum value of log~N(O~{\small VI}) is 14.04 atoms cm$^{-2}$ in superbubble N206 and the maximum value of log~N(O~{\small VI}) is 14.57 atoms cm$^{-2}$ in superbubble N70. The properties of O~{\small VI} in superbubbles reported here are tabulated in Table \ref{Tab3}. Comparing the abundance of O~{\small VI} for the superbubble and non-superbubble lines of sight, we find that the mean log~N(O~{\small VI}) for superbubble lines of sight is $\langle N_{SB}\rangle$ = 14.35 atoms cm$^{-2}$ while for the non-superbubble lines of sight this is $\langle N_{non-SB}\rangle$ = 14.19 atoms cm$^{-2}$. Thus, an excess O~{\small VI} abundance of about 40\% in superbubbles of the LMC is found in comparison to non-superbubble regions. Combining the \citet{Danforth06a} and \citet{Howk02a} data for superbubble and non-superbubble lines of sight (excluding Sk-67D05 sightline; see \citet{Howk02a} for details), the O~{\small VI} excess in superbubbles is about 46\%, which is comparable to our results. Thus, results reported here support and confirm that superbubbles do show higher O~{\small VI} abundance in comparison to the general halo absorption seen in other LMC lines of sight. Some of the non-superbubble lines of sight show an enhanced O~{\small VI} column density owing to local effects. \citet{Lehner07} compared superbubble and non-superbubble lines of sight for the LMC and found that some quiescent environments showed an enhanced O~{\small VI} abundance, sometimes even larger than that of superbubbles.

Even in a single superbubble, there are significant variations in the O~{\small VI} column densities. We have four observations each for the superbubbles 30 Doradus C and N11. In the case of 30 Doradus C, we find that the variation in N(O~{\small VI}) is more than a factor of 2 (from the minimum value of N(O~{\small VI}) to the maximum value). For N11, this variation is about a factor of 2.5. The variation in N(O~{\small VI}) for N70 is not much (for the three lines of sight included here) and this corroborates with the \citet{Danforth06a} data.

\section{Properties of O~{\small VI} in 30 Doradus}

The 30 Doradus region of the LMC is ideally suited to study the interaction between a high rate of star formation and the surrounding ISM. This region is dominated by the star forming cluster NGC 2070 that contains the very interesting star cluster R136 at the center. The proximity to 30 Doradus has allowed extensive research on the stellar content \citep{Parker92, Parker93a, Walborn97} and initial mass function (\citet{Andersen09} and references therein). 30 Doradus has also been investigated in detail in the infrared bands to study the dust properties \citep{Sturm00, Vermeij02, Meixner06}. \citet{Indebetouw09} studied the 30 Doradus in the mid-infrared wavelength band to determine the physical conditions of the ionized gas and found that the local effects of hot stars in 30 Doradus appear to dominate over any large-scale trend with distance from the central cluster R136.

The O~{\small VI} abundance in 30 Doradus is higher than in other regions of the LMC with the highest value of log~N(O~{\small VI}) being 14.56 atoms cm$^{-2}$ near the center of the cluster R136. Fig. \ref{Fig8} shows the change in the log~N(O~{\small VI}) values from the center of R136 up to an angular distance of 1 degree. We find an overall decrease in the O~{\small VI} abundance away from the center of R136. An interpretation of this plot could be that the processes involved in the formation of O~{\small VI} are likely to be associated with stellar radiation field but local effects cannot be neglected on a large angular scale. It should be noted that we do not find any correlation between O~{\small VI} absorption and H$_{\alpha}$ emission for LMC and thus, no relation between star formation is established on a larger scale.

As noticed earlier, log~N(O~{\small VI}) does not correlate with X-ray emission for the LMC as a whole but surprisingly there is a good correlation for the 30 Doradus lines of sight (Fig. \ref{Fig9}). The correlation is between log~N(O~{\small VI}) and X-ray emission from 30 Doradus considering all the lines of sight within 1 degree around R136 except for one sightline (SK-69D257) that has exceptionally high value of X-ray surface brightness. One of the reasons for such high X-ray emission from SK-69D257 is its proximity to a high mass X-ray binary LMC X-1 \citep{Points01}. The correlation in 30 Doradus suggests that O~{\small VI} observed in this region is present in the ISM gas surrounding the X-ray emitting plasma, which may be due to the compactness, high density and non-uniform structure of 30 Doradus. The correlation confirms that the X-ray emitting gas cools through temperatures where O~{\small VI} is being formed and supports the general consensus about the formation of O~{\small VI} by collisional ionization in the interface regions between cooler photo-ionized ISM gas and hot exterior ISM gas \citep{Slavin02, Indebetouw04}.

\section{Summary \& Conclusions}

We have presented O~{\small VI} column density measurements for the LMC using {\it FUSE} data for 70 lines of sight. This is the widest coverage of the LMC to date. The results reported here reveal significant variation in O~{\small VI} column densities over a very small angular scale thus confirming the patchiness of O~{\small VI} distribution in the LMC. The most important inferences drawn from this work are following:

\begin{enumerate}

\item{This survey probes 70 lines of sight with varying environmental conditions. We find strong O~{\small VI} absorption in the LMC that is not restricted to active regions. High O~{\small VI} abundance is present even in relatively inactive regions of the LMC.}

\item{There are significant variations in the velocity profiles of O~{\small VI} absorption. The O~{\small VI} absorption profile is broader than the MW absorption for many lines of sight but it should be noted that unambiguous separation of the MW and the LMC components is not possible for most of the lines of sight. This proves to be a significant hurdle in interpreting exact kinematics for O~{\small VI} in the LMC (for e.g. outflows from the LMC, etc.).}

\item{The maximum column density measured for the LMC is log~N(O~{\small VI}) = 14.57 atoms cm$^{-2}$ and minimum value is log~N(O~{\small VI}) = 13.72 atoms cm$^{-2}$. The mean value of O~{\small VI} column density is $<$log~N(O~{\small VI})$>$ = 14.23 atoms cm$^{-2}$, which is slightly lower than the earlier reported value. The median value of O~{\small VI} column density in the LMC comes to be 14.22 atoms cm$^{-2}$. The results corroborates with the previous finding that the distribution of O~{\small VI} in the LMC is patchy.}

\item{Despite the fact that the LMC has lower metallicity than the MW, the abundance of O~{\small VI} and properties of O~{\small VI} absorption are similar in both the galaxies. The mean of log N$_{\perp}$(O~{\small VI}) value for the MW is 14.29 atoms cm$^{-2}$ while the projected column density for the LMC, i.e., log N$_{\perp}$(O~{\small VI}) is 14.15 atoms cm$^{-2}$. A more extensive study for the MW suggests this value to be 14.21 atoms cm$^{-2}$. The SMC with even lower metallicity has higher O~{\small VI} abundance with mean log~N(O~{\small VI}) = 14.53 atoms cm$^{-2}$.}

\item{O~{\small VI} absorption in the LMC does not correlate with H$_{\alpha}$ (warm gas) or X-ray (hot gas) emission but we find a good correlation between O~{\small VI} absorption and X-ray emission in the 30 Doradus region. It is also seen that the O~{\small VI} absorption is decreasing with increasing angular distance from the star cluster R136 suggesting some correlation with stellar radiation field.}

\item{The observations reported cover 10 superbubbles of the LMC and for 5 superbubbles (30 Doradus C, N158, N51, N11 and N57) O~{\small VI} absorption is reported for the first time. Superbubbles are O~{\small VI} rich and have 40\% excess compared to non-superbubble lines of sight.}

\end{enumerate}

\section*{Acknowledgments}

AP and NVS acknowledge fellowships from the Department of Science and Technology, New Delhi. This research has made use of {\it Far Ultraviolet Spectroscopic Explorer} data. {\it FUSE} was operated by Johns Hopkins University for NASA. We acknowledge the use of Multimission Archive at Space Telescope Science Institute (MAST) and NASA Astrophysics Data System (ADS).

\clearpage

\begin{figure}
\includegraphics{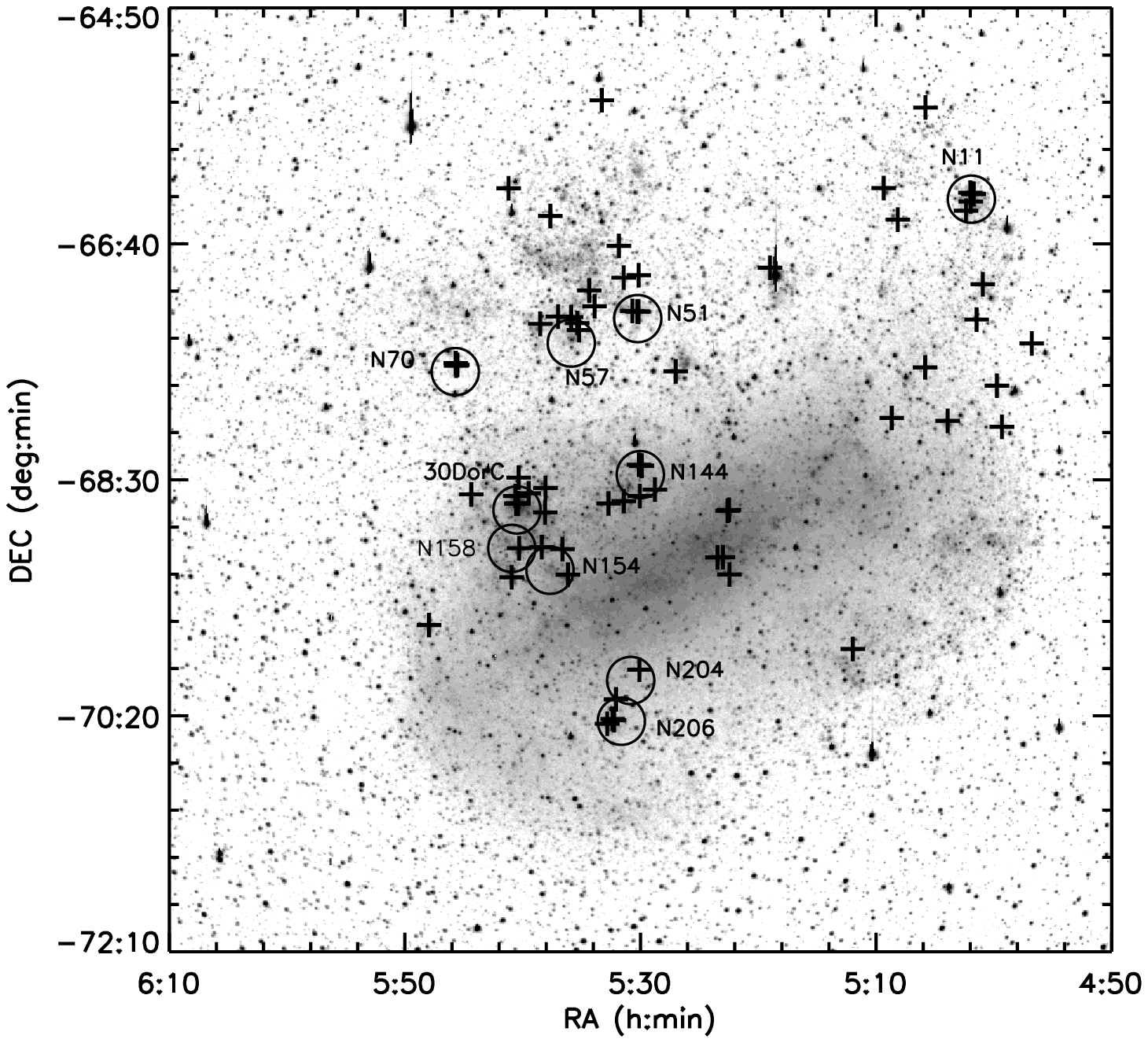}
\vspace{3.5cm}
\caption{R-band image of the LMC \citep{Bothun88} showing the targets towards which O~{\small VI} absorption have been studied. Superbubbles have been shown by circles.}
\label{Fig1}
\end{figure}

\clearpage

\begin{figure}
\includegraphics{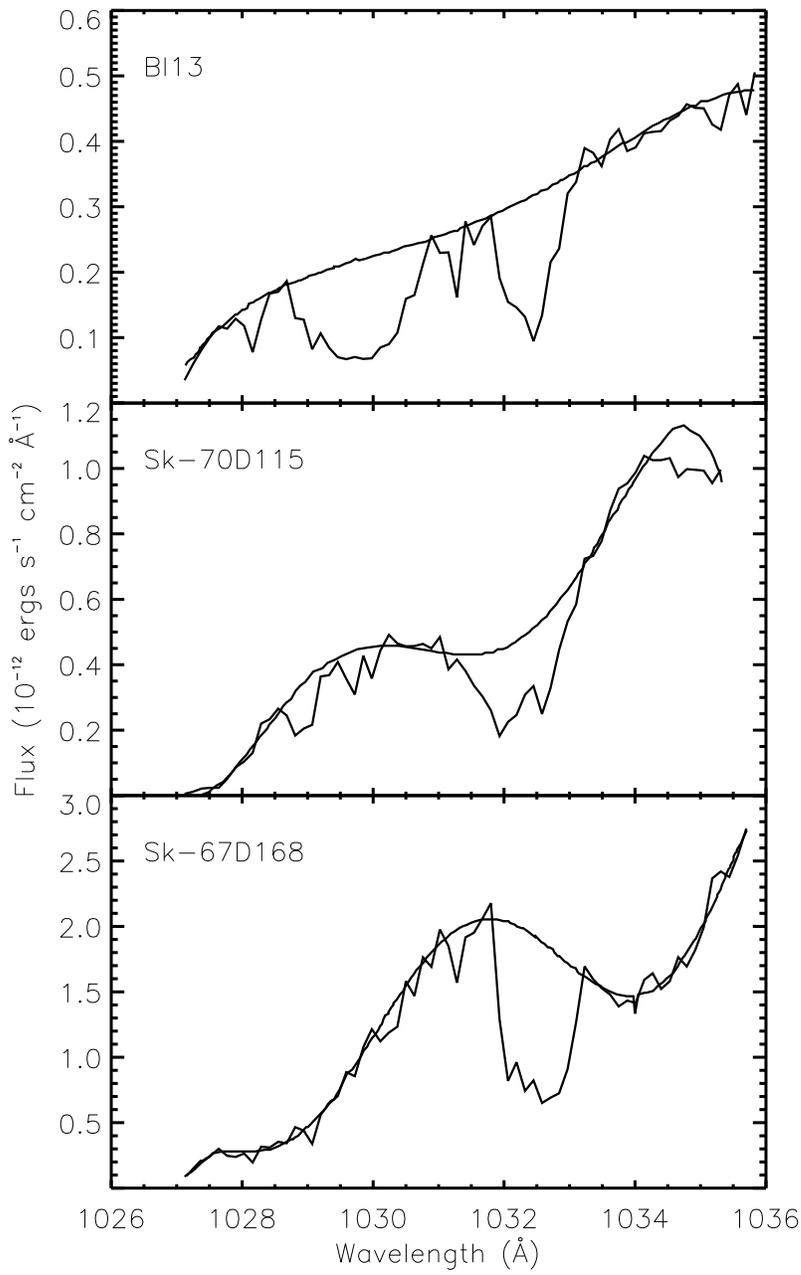}
\vspace{3.5cm}
\caption{Examples for continuum fitting for three sightlines in the region near O~{\small VI} absorption at 1031.92 \AA. The top to bottom panels show the increasing complexity of fitting the stellar continuum.}
\label{Fig2}
\end{figure}

\clearpage

\begin{figure}
\subfigure[Normalized O~{\small VI} absorption profiles for the 70 lines of sight. Table \ref{Tab1} gives the details of each sightline.]
 {\label{Fig3a}\includegraphics{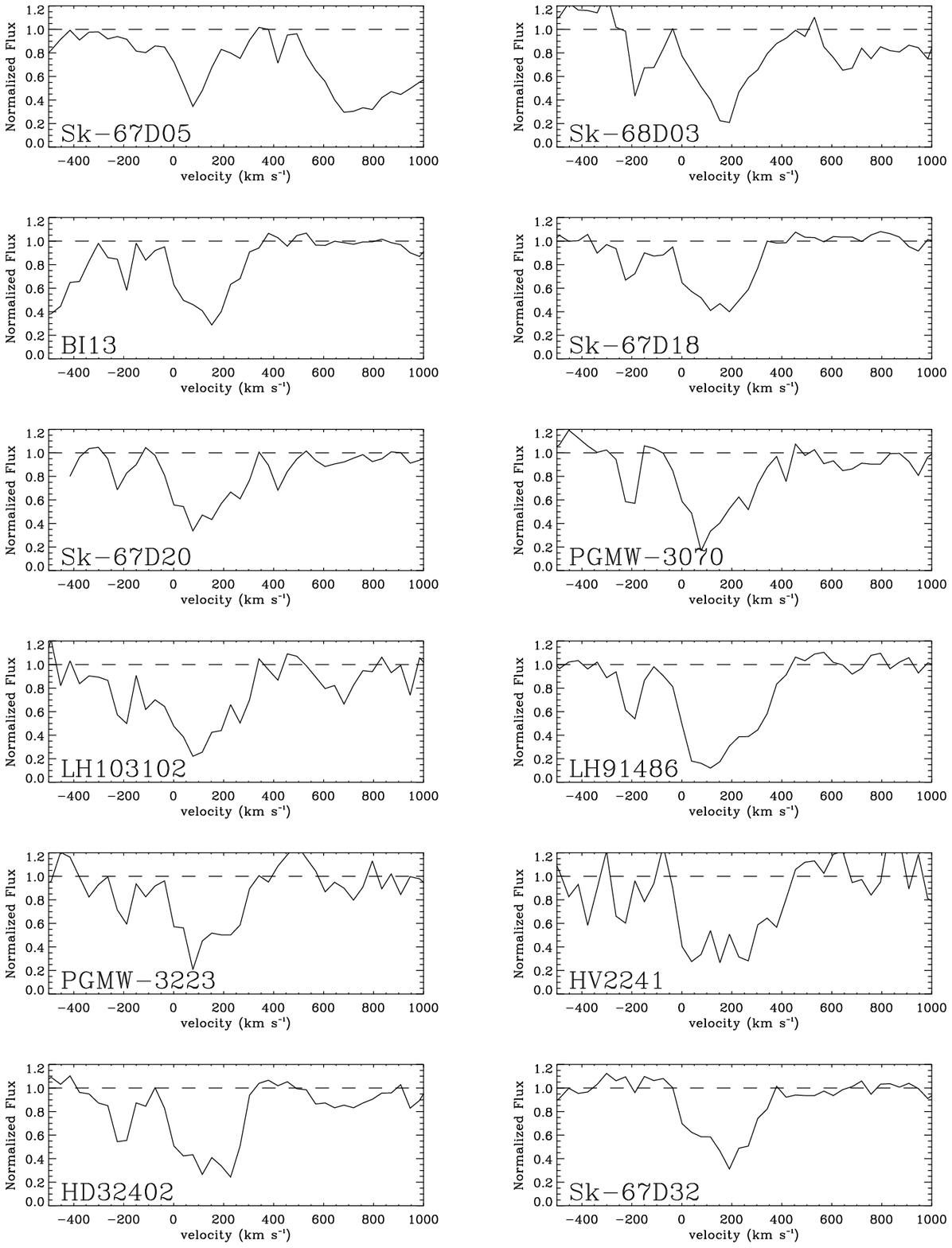}}\clearpage
\captcont{} 
\end{figure}

\clearpage

\begin{figure}
 \subfigure[Same as Fig. \ref{Fig3a}]
 {\label{Fig3b}\includegraphics{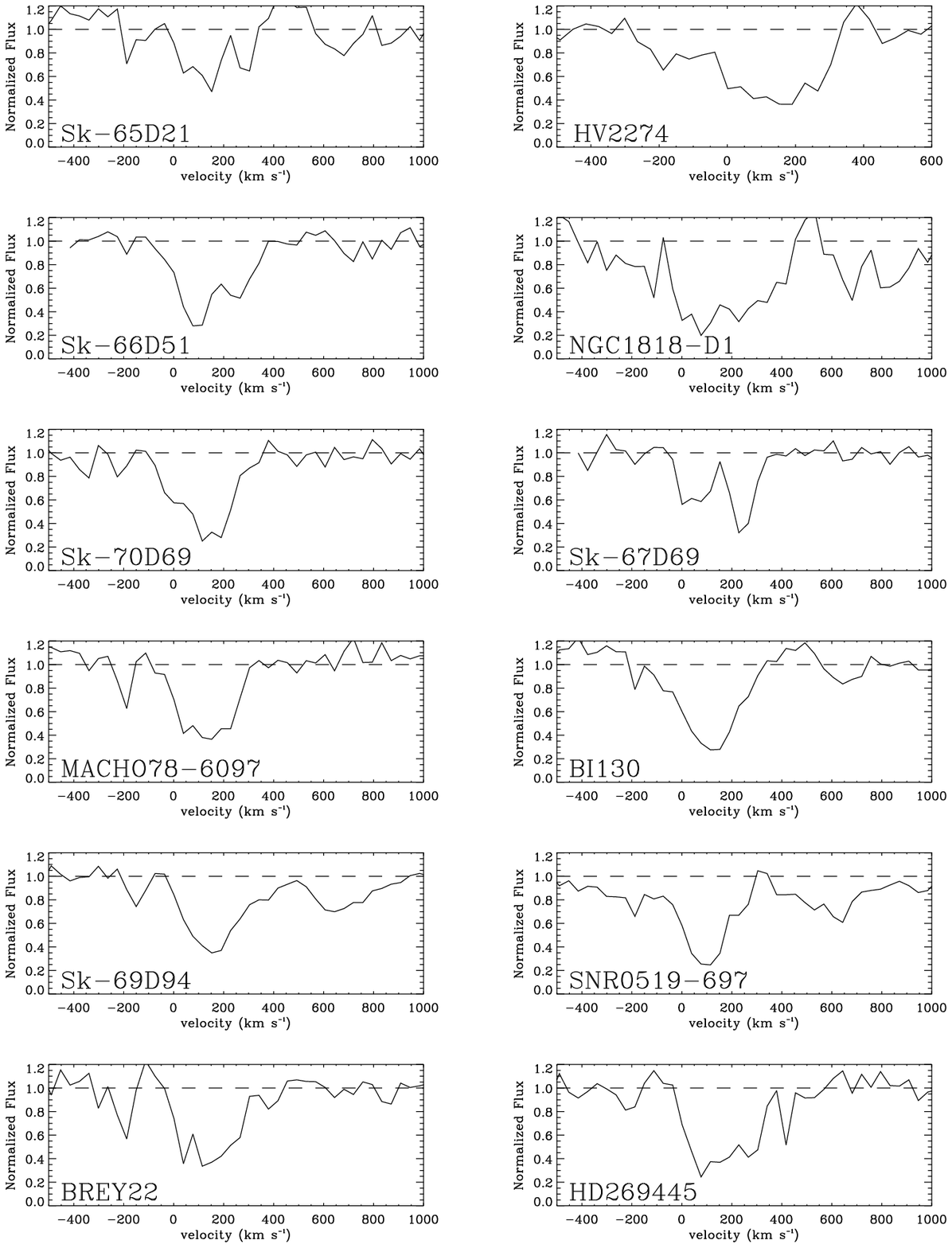}}
\captcont{} 
\end{figure}

\clearpage

\begin{figure}
 \subfigure[Same as Fig. \ref{Fig3a}]
 {\label{Fig3c}\includegraphics{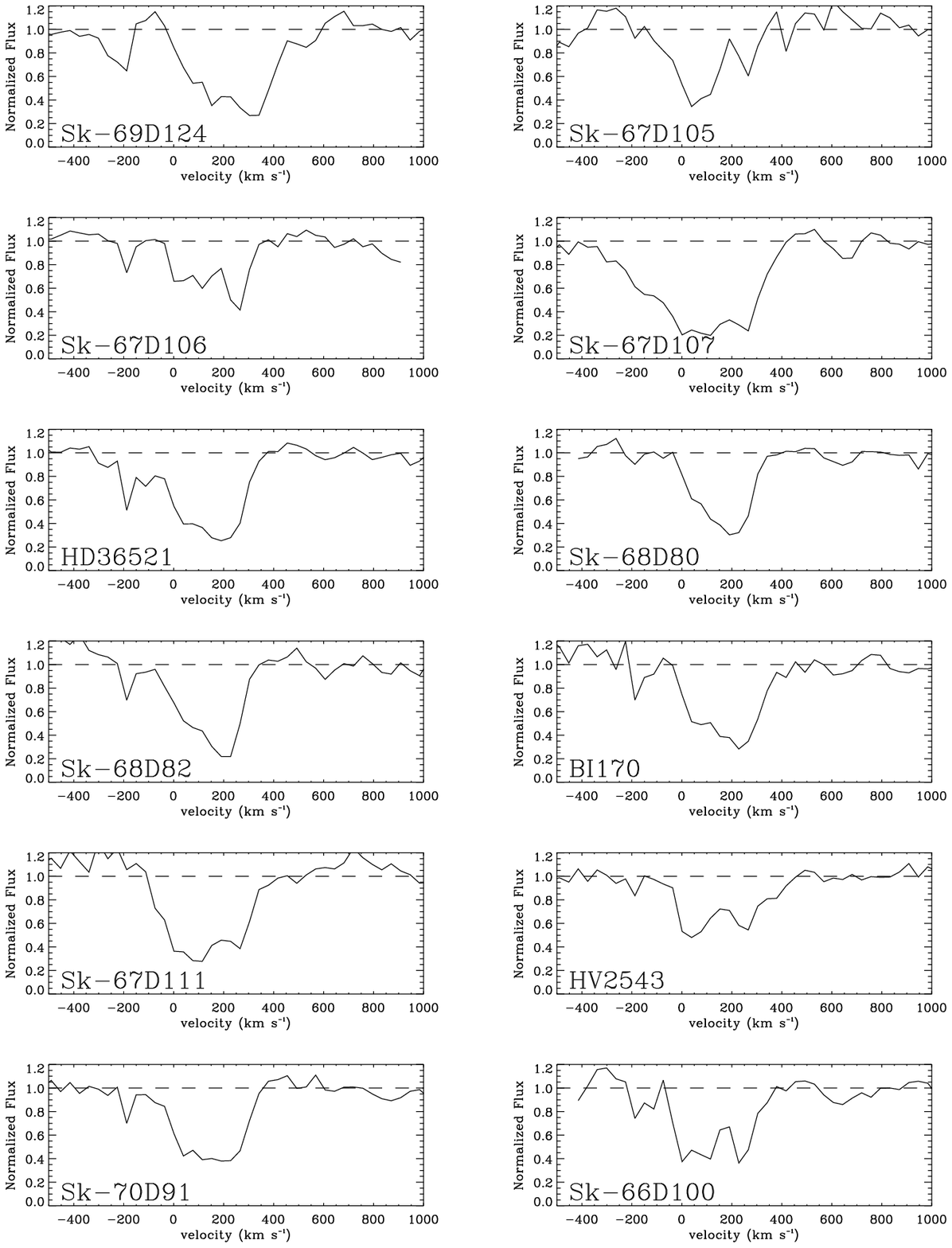}}
\captcont{} 
\end{figure}

\clearpage

\begin{figure}
 \subfigure[Same as Fig. \ref{Fig3a}]
 {\label{Fig3d}\includegraphics{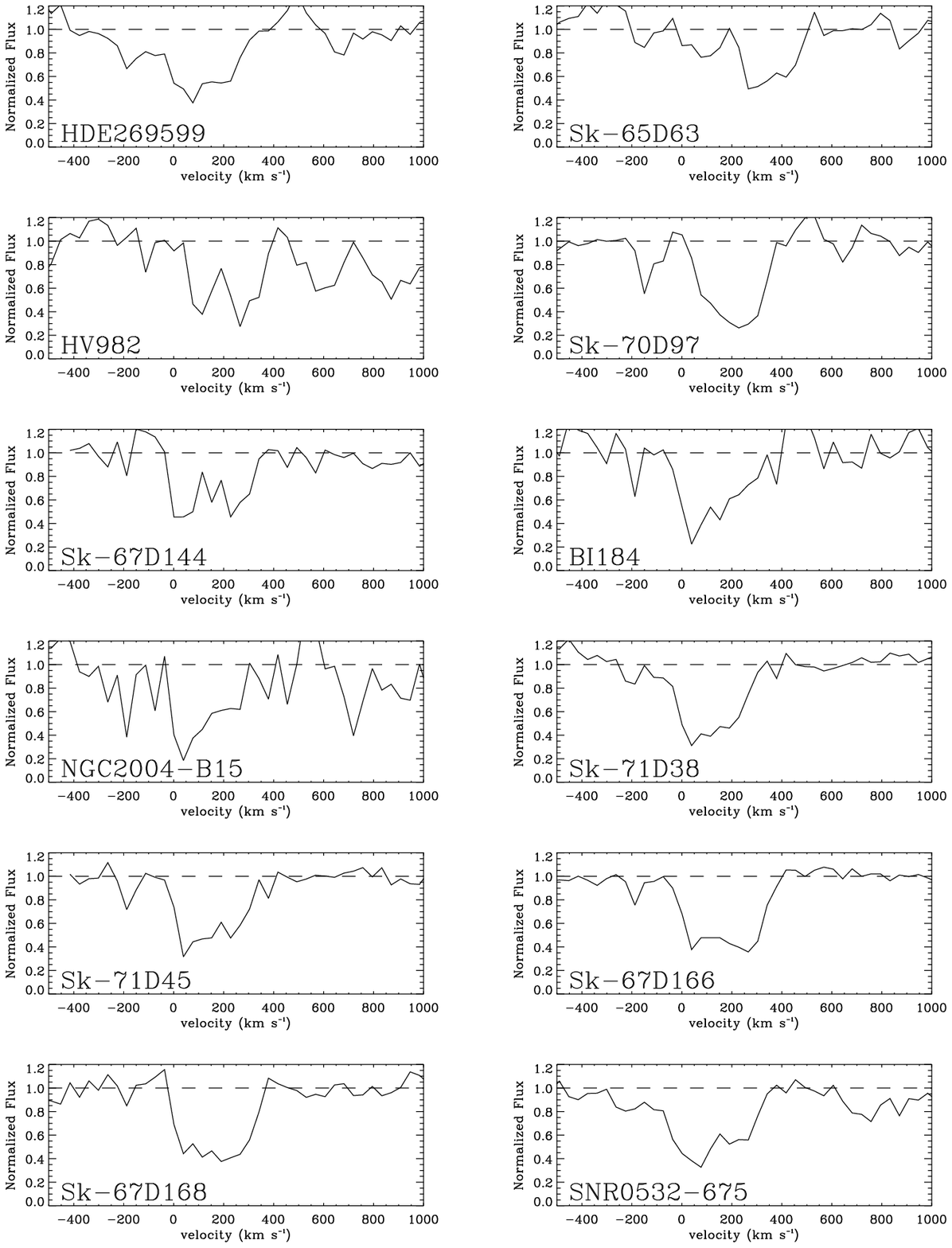}}
\captcont{} 
\end{figure}

\clearpage

\begin{figure}
 \subfigure[Same as Fig. \ref{Fig3a}]
 {\label{Fig3e}\includegraphics{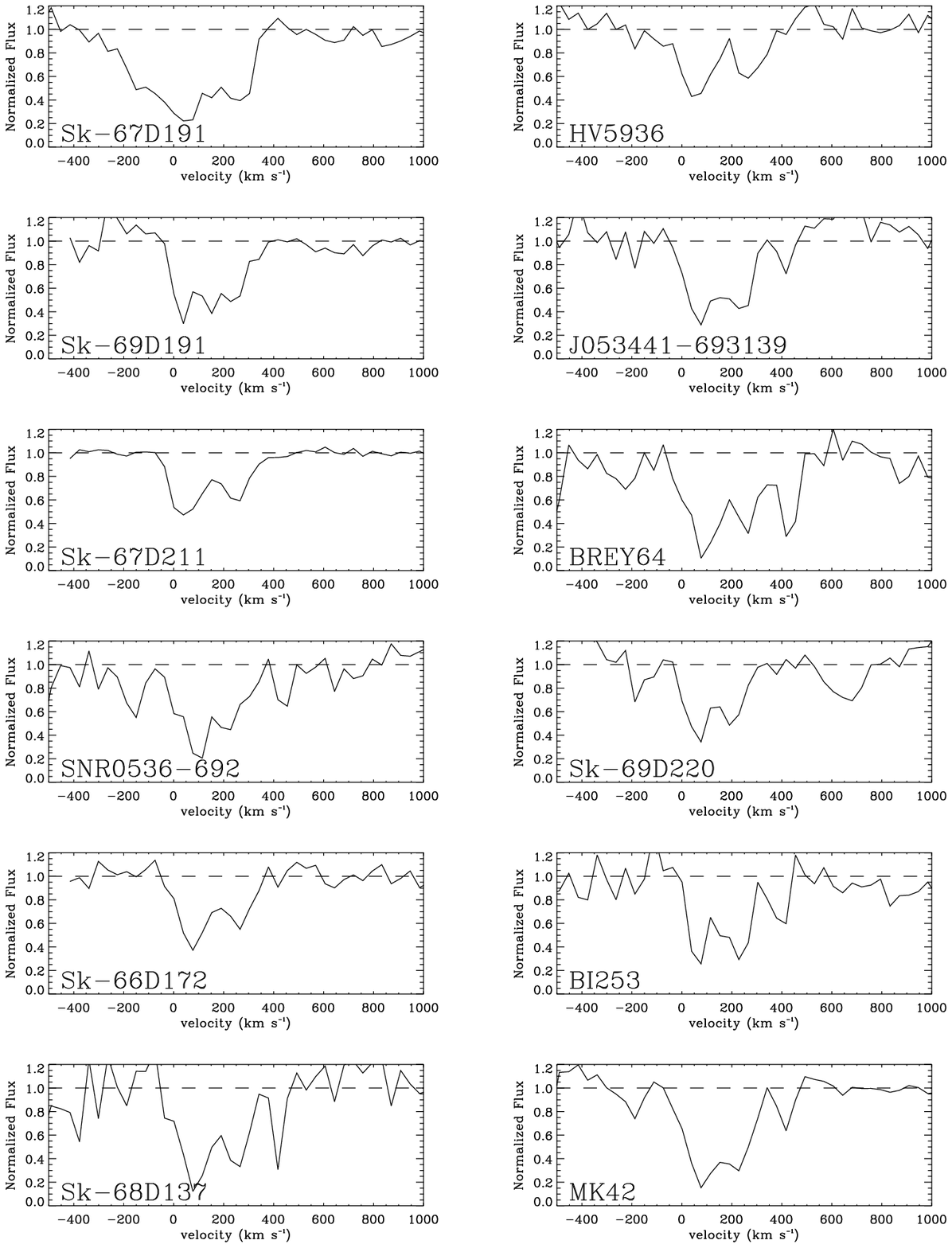}}
\captcont{}
\end{figure}

\clearpage

\begin{figure}
 \subfigure[Same as Fig. \ref{Fig3a}]
 {\label{Fig3f}\includegraphics{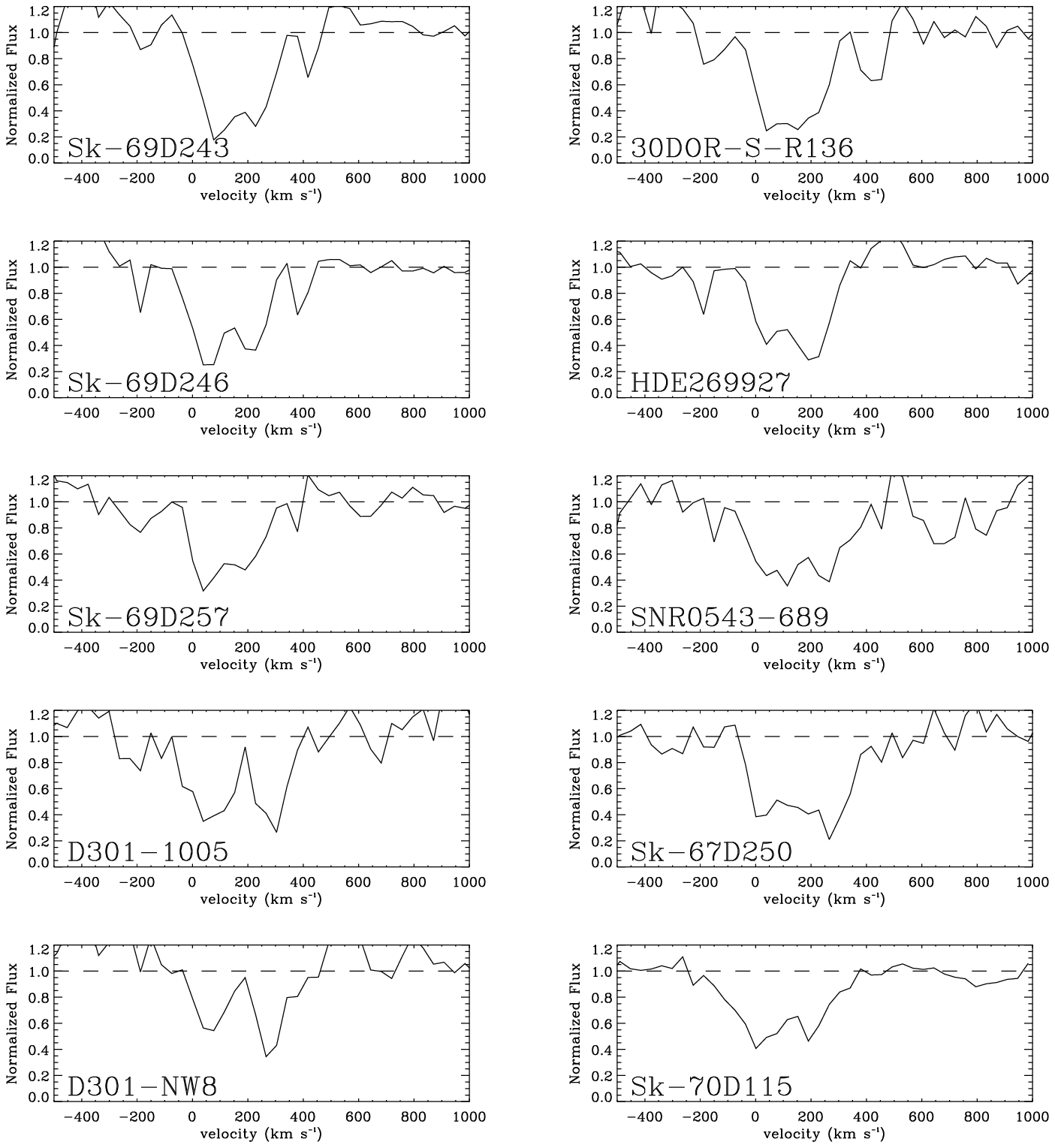}}
\caption*{}
  \label{Fig3}
\end{figure}

\clearpage

\begin{figure}
\includegraphics{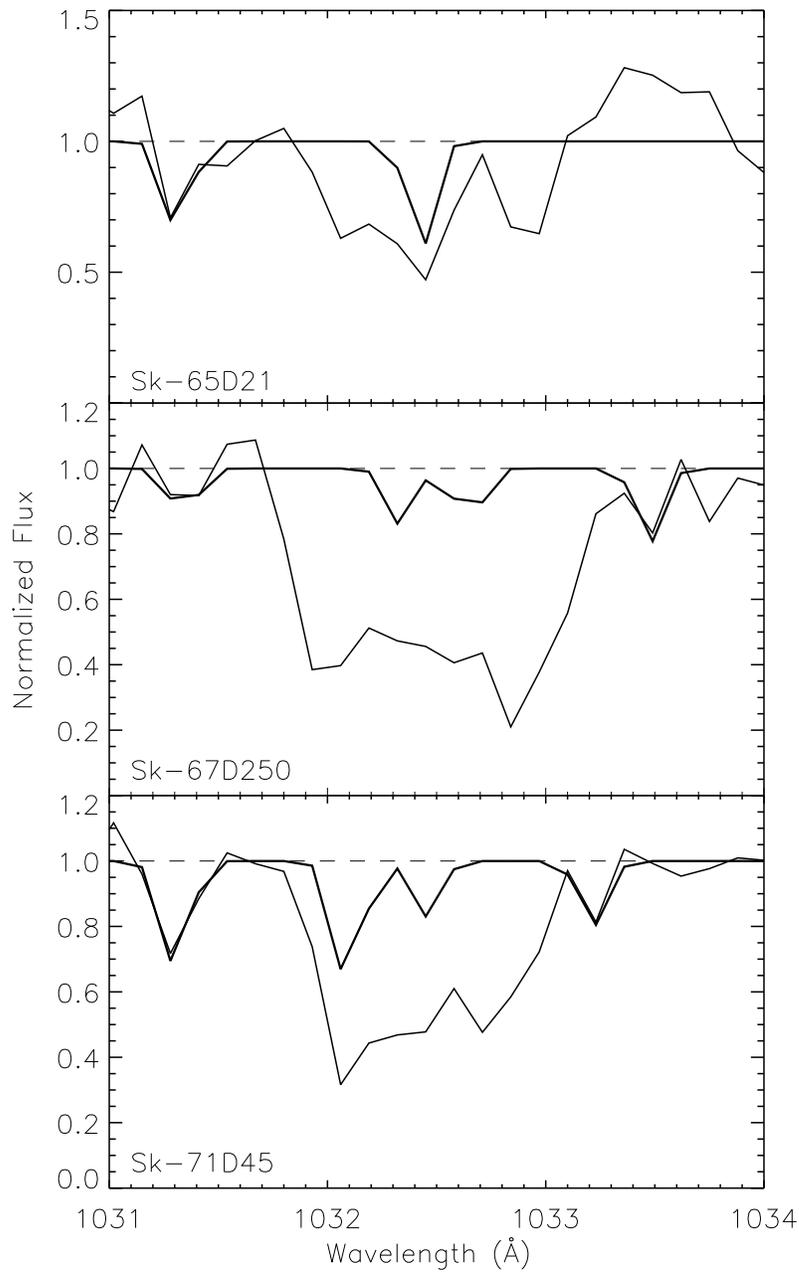}
\vspace{3.5cm}
\caption{Model for H$_2$ absorption overplotted on the normalized O~{\small VI} absorption profiles for 3 lines of sight. For Sk-65D21, H$_2$ absorption does not contaminate the LMC O~{\small VI} absorption, while Sk-71D45 shows the maximum contamination. Sk-67D250 shows moderate contamination.}
\label{Fig4}
\end{figure}

\clearpage

\begin{figure}
\includegraphics{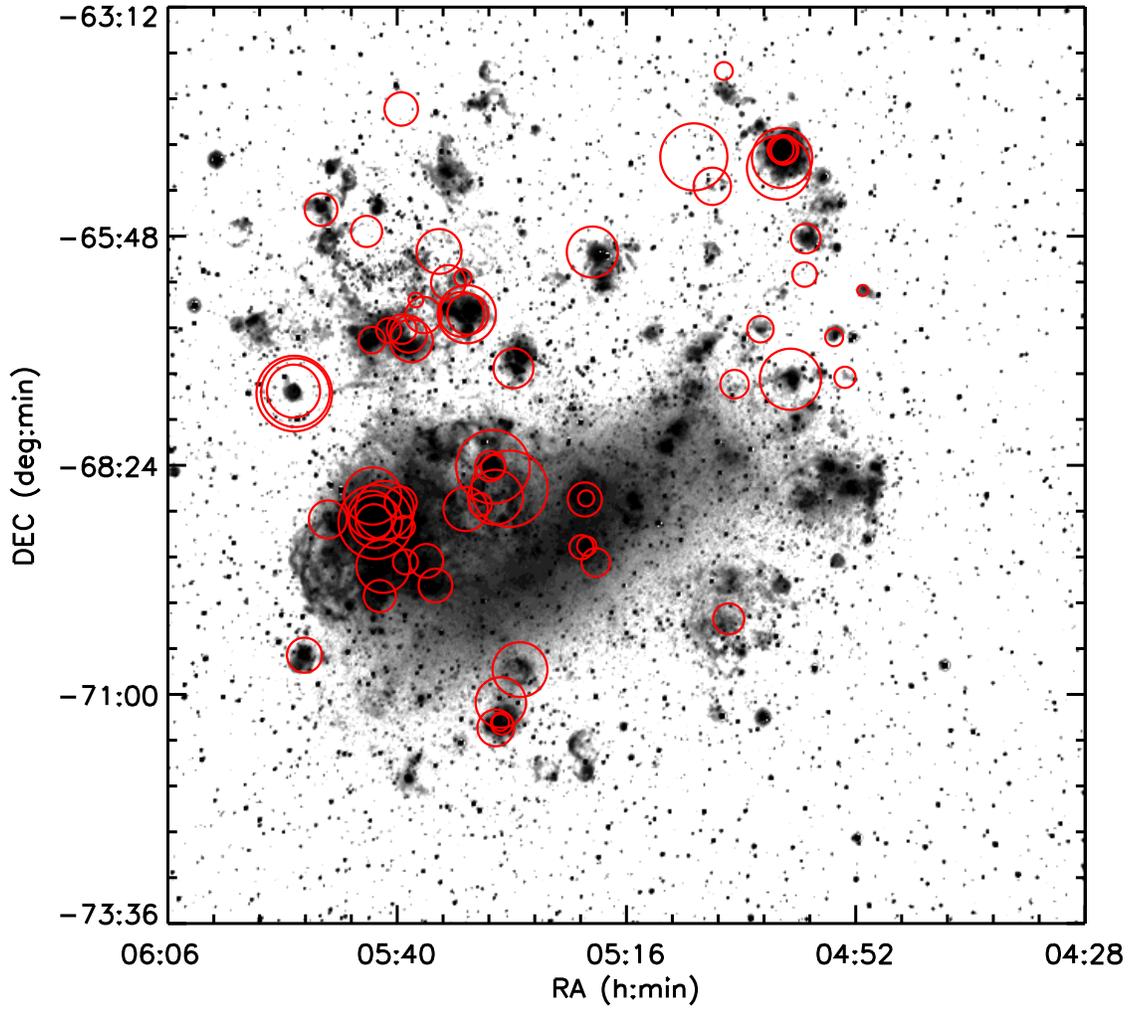}
\vspace{3.5cm}
\caption{H$_{\alpha}$ image of the LMC \citep{Gaustad01} with circles showing O~{\small VI} absorption around the 70 targets. The diameter of the circle is linearly proportional to the column density of O~{\small VI} at LMC velocities.}
\label{Fig5}
\end{figure}

\clearpage

\clearpage

\begin{figure}
\includegraphics{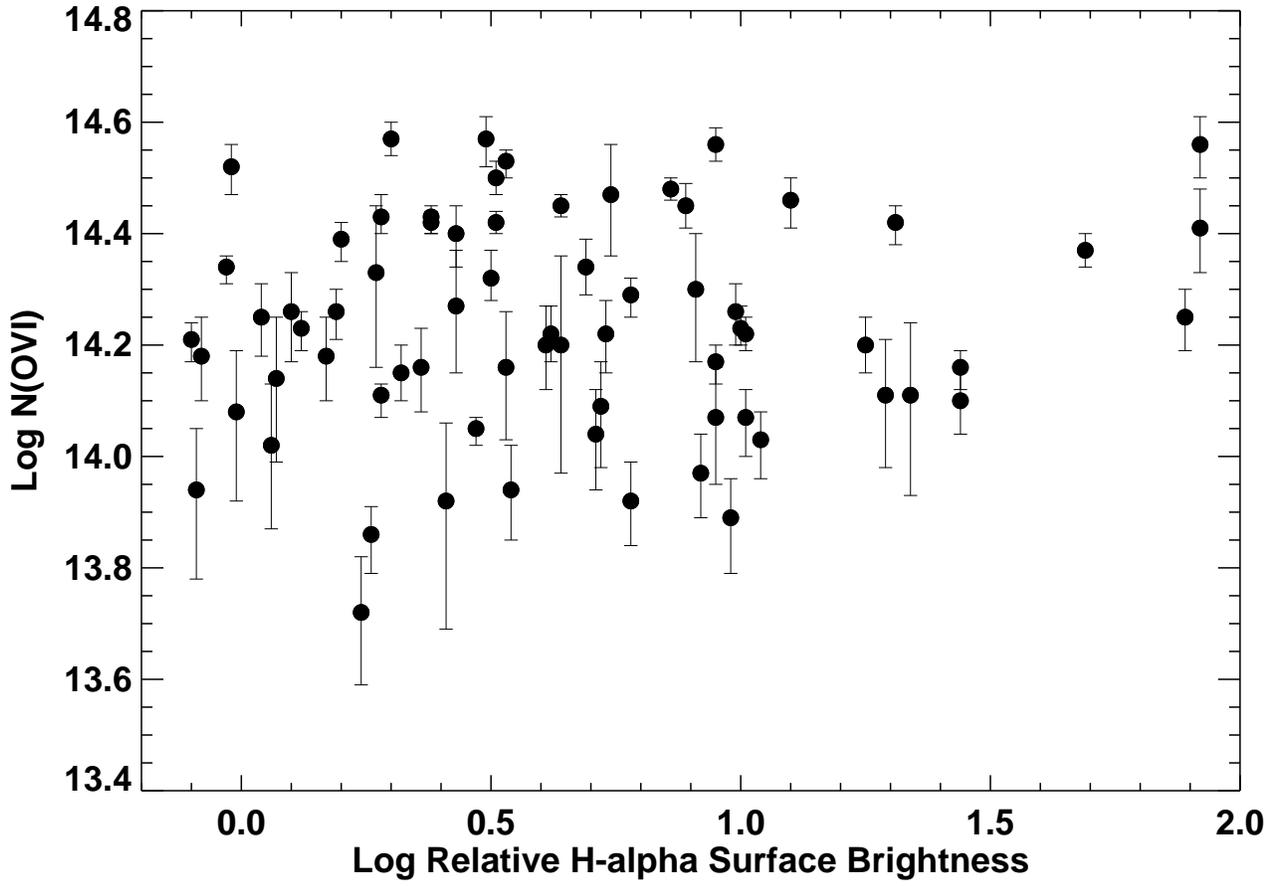}
\vspace{3.5cm}
\caption{O~{\small VI} column density (log~N(O~{\small VI})) vs. log relative H$_{\alpha}$ surface brightness for all the 70 targets. H$_{\alpha}$ surface brightness is from \citet{Gaustad01}.}
\label{Fig6}
\end{figure}

\clearpage

\begin{figure}
\includegraphics{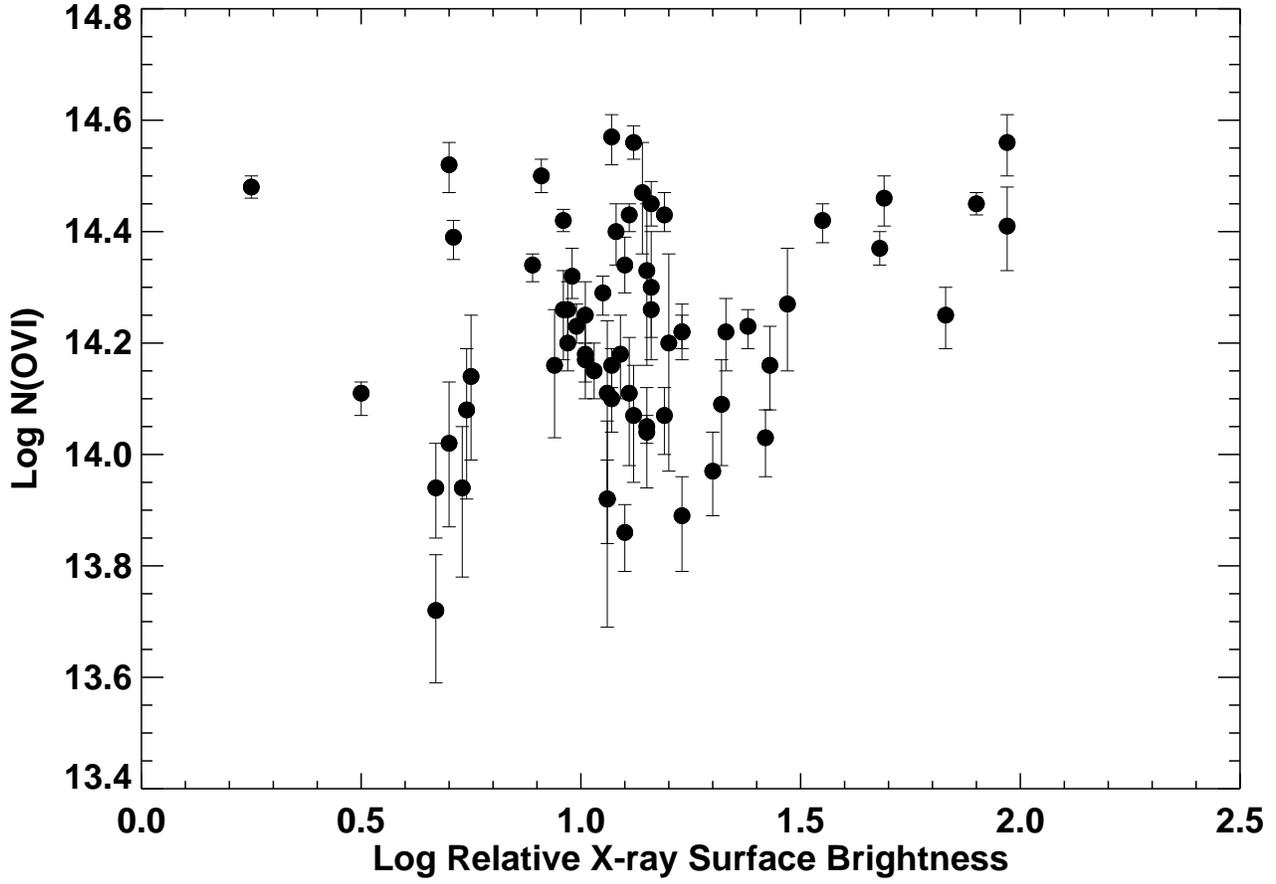}
\vspace{3.5cm}
\caption{O~{\small VI} column density (log~N(O~{\small VI})) vs. log relative X-ray surface brightness for 65 targets. The X-ray surface brightness is taken from the {\it ROSAT} PSPC mosaic image (energy range 0.5--2.0 keV) \citep{Snowden94}. 4 targets (Sk-67D250, D301-1005, D301-NW8 and Sk-65D63) have not been covered by X-ray observations and one of the target Sk-69D257 has been excluded because it has extremely high X-ray emission.}
\label{Fig7}
\end{figure}

\clearpage

\begin{figure}
\includegraphics{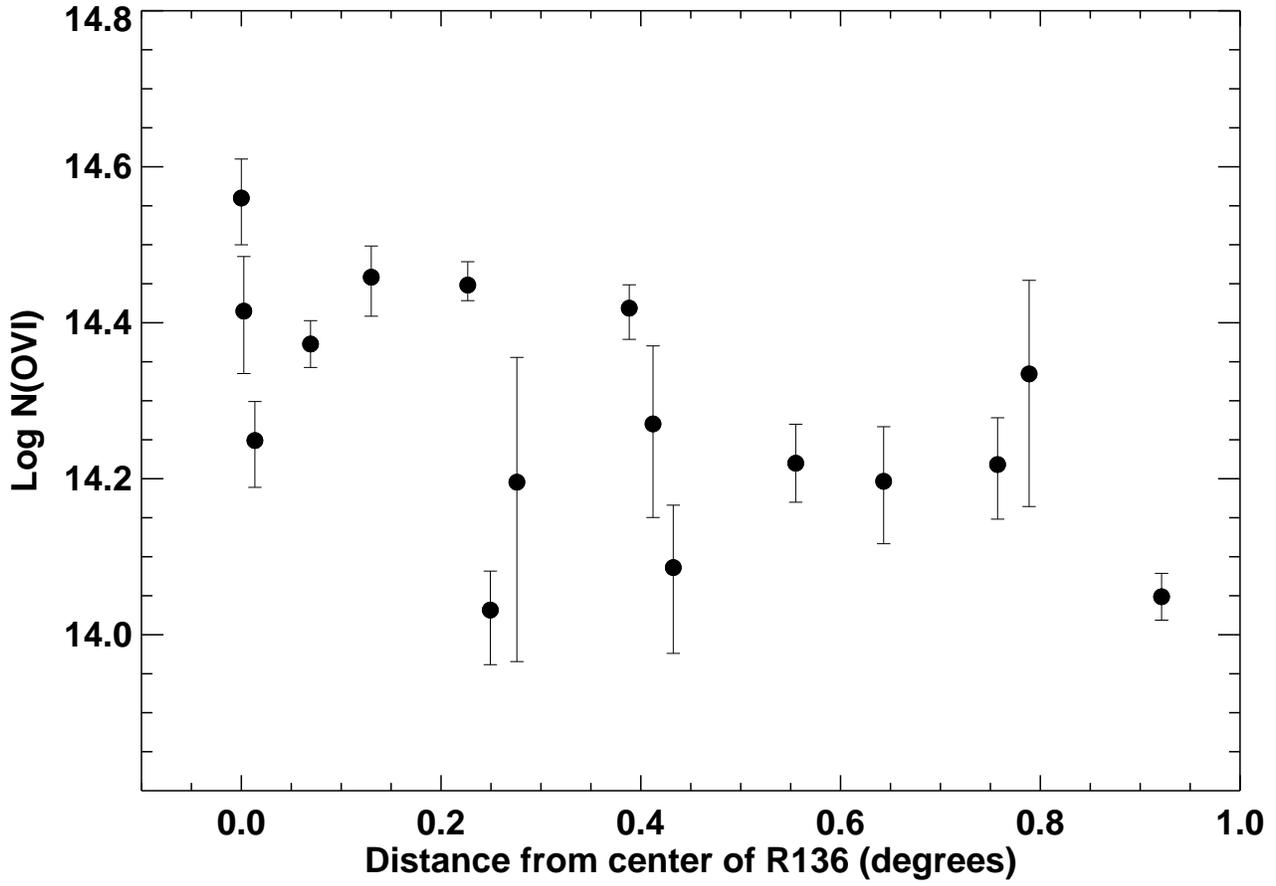}
\vspace{3.5cm}
\caption{Variation of O~{\small VI} column density (log~N(O~{\small VI})) with increasing distance from the the centre of the star cluster R136 located in the 30 Doradus region of the LMC. Note that the variation is plotted within an angular distance of 1 degree from the the centre of R136.}
\label{Fig8}
\end{figure}

\clearpage

\begin{figure}
\includegraphics{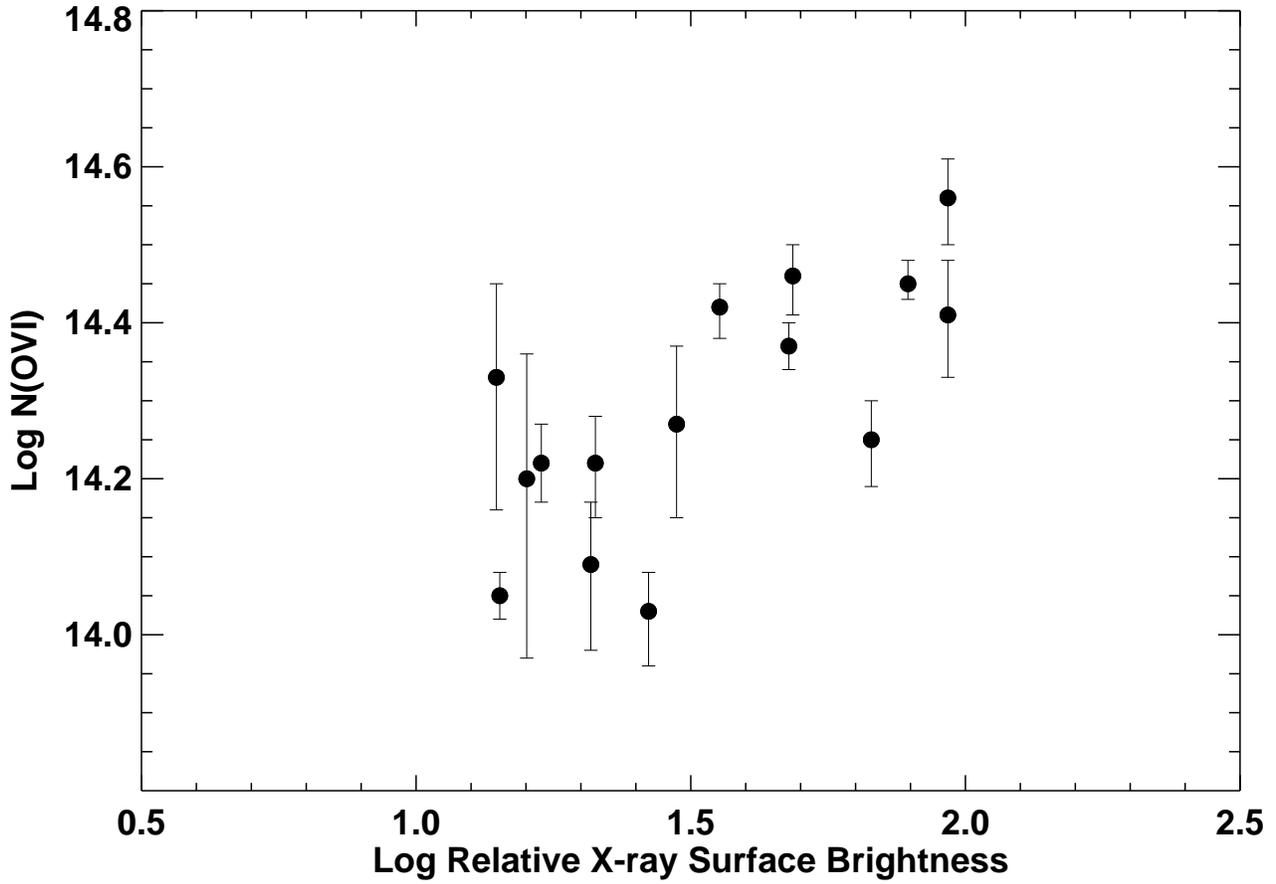}
\vspace{3.5cm}
\caption{O~{\small VI} column density (log~N(O~{\small VI})) vs. X-ray luminosity for the 30 Doradus region of the LMC. The X-ray surface brightness is taken from the {\it ROSAT} PSPC mosaic image (energy range 0.5--2.0 keV) \citep{Snowden94}. A linear correlation is present which is not seen when all the targets of LMC are included (Fig. \ref{Fig7}) and for other individual regions in the LMC. The plot includes all lines of sight from Fig. \ref{Fig8} except for 1 sightline (SK-69D257) as it shows an exceptionally high X-ray emission due to its proximity to X-ray binary LMC-X1 (Points et al. 2001).}
\label{Fig9}
\end{figure}

\clearpage
\onecolumn
\scriptsize

i%\begin{table}
\begin{longtable}{lccccccc}
\caption{Log of {\it FUSE} observations for the 70 targets in the LMC.}
\label{Tab1}\\
%\begin{tabular}{l c c c c c c c}

\hline \\[-2ex]
\multicolumn{1}{l}{{\it FUSE} ID}  &
\multicolumn{1}{c}{Object name}    &
\multicolumn{1}{c}{{\it FUSE} aperture}  &
\multicolumn{1}{c}{RA}             &
\multicolumn{1}{c}{Dec}            &
\multicolumn{1}{c}{Spectral Type}          &
\multicolumn{1}{c}{V mag.} &
\multicolumn{1}{c}{Reference}\\[0.5ex] \hline \\[-1.8ex]
\endfirsthead

\multicolumn{8}{c}{{\tablename} \thetable{} -- Continued} \\[0.5ex]
\hline \hline \\[-2ex]
\multicolumn{1}{l}{{\it FUSE} ID}  &
\multicolumn{1}{c}{Object name}    &
\multicolumn{1}{c}{{\it FUSE} aperture}  &
\multicolumn{1}{c}{RA}             &
\multicolumn{1}{c}{Dec}            &
\multicolumn{1}{c}{Spectral Type}          &
\multicolumn{1}{c}{V mag.} &
\multicolumn{1}{c}{Reference}\\[0.5ex] \hline \\[-1.8ex]
\endhead

\multicolumn{8}{l}{{Continued on Next Page\ldots}} \\
\endfoot

%\\[-1.8ex] \hline \hline
\endlastfoot
%\hline
%{\it FUSE} ID  &Object name  &{\it FUSE} aperture  &RA  &Dec  &Spectral Type  &V mag.  &Reference\\
%\hline
P1030703        &Sk-67D05       &LWRS           &04 50 18.96    &-67 39 37.9    &O9.7Ib                 &11.34  &1\\
A0490402        &Sk-68D03       &LWRS           &04 52 15.56    &-68 24 26.9    &O9I                    &13.13  &2\\
E5111901        &BI13           &LWRS           &04 53 06.48    &-68 03 23.1    &O6.5V                  &13.75  &3\\
D0980801        &Sk-67D18       &MDRS           &04 55 14.90    &-67 11 24.5    &O6-7n-nm+WN5-6A        &11.95  &1\\
P1174401        &Sk-67D20       &LWRS           &04 55 31.50    &-67 30 01.0    &WN4b                   &13.86  &4\\
B0100201        &PGMW-3070      &MDRS           &04 56 43.25    &-66 25 02.0    &O6V                    &12.75  &5\\
D0300302        &LH103102       &MDRS           &04 56 45.40    &-66 24 45.9    &O7Vz                   &13.55  &5\\
D0300101        &LH91486        &MDRS           &04 56 55.58    &-66 28 58.0    &O6.5Vz                 &14.2   &5\\
B0100701        &PGMW-3223      &MDRS           &04 57 00.80    &-66 24 25.3    &O8.5IV                 &12.95  &5\\
F9270604        &HV2241         &LWRS           &04 57 15.83    &-66 33 54.8    &O7-III                 &13.5   &5\\
P1174901        &HD32402        &LWRS           &04 57 24.19    &-68 23 57.2    &WC                     &       &3\\
Z9050801        &Sk-67D32      &LWRS           &04 59 52.30    &-67 56 55.0    &WN                     &14.48  &3\\
P1030901        &Sk-65D21       &LWRS           &05 01 22.33    &-65 41 48.1    &O9.7Iab                &12.02  &6\\
B0770201        &HV2274         &LWRS           &05 02 40.96    &-68 24 21.3    &B0-B2III-I             &14.2   &5\\
P1174501        &Sk-66D51       &LWRS           &05 03 10.20    &-66 40 54.0    &WN8h                   &12.71  &4\\
D1380201        &NGC1818-D1     &LWRS           &05 04 32.56    &-66 24 51.0    &B0-B2V-IV              &14.93  &5\\
P1172101        &Sk-70D69       &LWRS           &05 05 18.73    &-70 25 49.8    &O5V                    &13.94  &6\\
P1171703        &Sk-67D69       &LWRS           &05 14 20.16    &-67 08 03.5    &O4III(f)               &13.09  &7\\
C1030101        &MACHO78-6097   &LWRS           &05 18 04.70    &-69 48 19.0    &B0-B2V-IV              &14.4   &5\\
E5112902        &BI130          &LWRS           &05 18 06.06    &-69 14 34.5    &O8.5V((f))             &12.53  &3\\
Z9050201        &Sk-69D94      &LWRS           &05 18 14.53    &-69 15 01.0    &B0-B2III-I             &9.72   &5\\
D9044701        &SNR0519-697    &LWRS           &05 18 44.20    &-69 39 12.4    &Supernova-remnant      &       &\\
D0880104        &BREY22         &MDRS           &05 19 16.40    &-69 39 19.5    &WC/O9.5Ib              &12.3   &1\\
B0270101        &HD269445       &LWRS           &05 22 59.87    &-68 01 46.6    &WN                     &11.45  &1\\
P1173602        &Sk-69D124      &LWRS           &05 25 18.37    &-69 03 11.1    &O9Ib                   &12.81  &2\\
D1530104        &Sk-67D105      &LWRS           &05 26 06.37    &-67 10 57.6    &Of                     &12.42  &5\\
A1110101        &Sk-67D106      &MDRS           &05 26 15.20    &-67 29 58.3    &B0-B2III-I             &11.78  &8\\
A1110201        &Sk-67D107      &MDRS           &05 26 20.67    &-67 29 55.4    &B0-B2III-I             &12.5   &8\\
C1510101        &HD36521        &LWRS           &05 26 30.32    &-68 50 25.4    &WC                     &12.42  &9\\
P1031402        &Sk-68D80       &LWRS           &05 26 30.43    &-68 50 26.6    &WC4+OB                 &12.40  &10\\
P2030104        &Sk-68D82       &LWRS           &05 26 45.31    &-68 49 52.8    &WN5?b+(B3I)            &9.86   &5\\
P1173701        &BI170          &LWRS           &05 26 47.79    &-69 06 11.7    &O9.5Ib                 &13.09  &11\\
C1550201        &Sk-67D111      &LWRS           &05 26 47.95    &-67 29 29.9    &O6:Iafpe               &12.57  &11\\
F9270701        &HV2543         &LWRS           &05 27 27.40    &-67 11 55.4    &O8+O9                  &12.92  &5\\
P1172501        &Sk-70D91       &LWRS           &05 27 33.74    &-70 36 48.3    &O6.5V                  &12.78  &11\\
P1172303        &Sk-66D100      &LWRS           &05 27 45.59    &-66 55 15.0    &O6II(f)                &13.26  &6\\
E9570301        &HDE269599      &LWRS           &05 28 22.68    &-69 08 32.2    &Be                     &10.18  &5\\
M1142001        &Sk-65D63       &LWRS           &05 28 39.50    &-65 39 01.1    &O9.7I                  &12.56  &12\\
C1030201        &HV982          &LWRS           &05 29 52.50    &-69 09 22.0    &B0-B2V-IV              &14.6   &5\\
E5113602        &Sk-70D97       &LWRS           &05 30 11.35    &-70 51 42.2    &O9III                  &13.33  &5\\
P1175001        &Sk-67D144      &LWRS           &05 30 12.22    &-67 26 08.4    &WC4+OB                 &13.6   &13\\
P2170301        &BI184          &LWRS           &05 30 30.60    &-71 02 31.3    &B0-B2V-IV              &13.84  &3\\
D1380301        &NGC2004-B15    &MDRS           &05 30 36.58    &-67 17 42.3    &B0-B2V-IV              &14.18  &5\\
Z9050601        &Sk-71D38      &LWRS           &05 30 38.77    &-71 01 47.9    &WC                     &13.1   &5\\
P1031502        &Sk-71D45       &LWRS           &05 31 15.55    &-71 04 08.9    &O4-5III(f)             &11.51  &1\\
A1330123        &Sk-67D166      &LWRS           &05 31 44.31    &-67 38 00.6    &O4If+                  &12.27  &1\\
B0860901        &Sk-67D168      &MDRS           &05 31 52.10    &-67 34 20.8    &O8Iaf                  &12.08  &12\\
D9042002        &SNR0532-675    &LWRS           &05 32 23.00    &-67 31 02.0    &Supernova-remnant      &       &\\
P1173101        &Sk-67D191      &LWRS           &05 33 34.12    &-67 30 19.6    &O8V                    &13.46  &2\\
B0770302        &HV5936         &LWRS           &05 33 39.00    &-66 37 39.8    &B0-B2V-IV              &14.8   &5\\
P1175101        &Sk-69D191      &LWRS           &05 34 19.39    &-69 45 10.0    &WC4                    &13.35  &13\\
F3210301        &J053441-693139 &LWRS           &05 34 41.30    &-69 31 39.0    &O2-O3.5If*             &13.7   &14\\
P1171603        &Sk-67D211      &LWRS           &05 35 13.92    &-67 33 27.0    &O2III(f)*              &12.28  &14\\
Z9051001        &BREY64         &LWRS           &05 35 54.45    &-68 59 07.4    &WN9h                   &13.21  &3\\
D9043101        &SNR0536-692    &LWRS           &05 36 07.70    &-69 11 52.6    &Supernova-remnant      &       &\\
Z9050701        &Sk-69D220     &LWRS           &05 36 43.83    &-69 29 47.4    &WNorOiafpe             &10.58  &1\\
P1172201        &Sk-66D172      &LWRS           &05 37 05.56    &-66 21 35.7    &O2III(f)*+OB           &13.13  &14\\
C0020601        &BI253          &LWRS           &05 37 34.49    &-69 01 09.8    &Of                     &13.76  &3\\
E5114201        &Sk-68D137      &LWRS           &05 38 24.77    &-68 52 32.8    &O3IIIf                 &13.29  &3\\
P1171803        &MK42           &LWRS           &05 38 42.10    &-69 05 54.7    &O3If/WN6-A             &10.96  &15\\
P1031706        &Sk-69D243      &LWRS           &05 38 42.57    &-69 06 03.2    &WN5+OB                 &9.5    &16\\
F9140101        &30DOR-S-R136   &LWRS           &05 38 51.70    &-69 06 00.0    &HII-region             &       &\\
P1031802        &Sk-69D246      &LWRS           &05 38 53.50    &-69 02 00.7    &WN                     &11.16  &4\\
P1174601        &HDE269927      &LWRS           &05 38 58.25    &-69 29 19.1    &WN                     &12.63  &4\\
P2170101        &Sk-69D257      &LWRS           &05 39 58.91    &-69 44 03.2    &O9II                   &12.53  &5\\
D9042801        &SNR0543-689    &LWRS           &05 43 07.20    &-68 58 52.0    &Supernova-remnant      &       &\\
D0981401        &D301-1005      &MDRS           &05 43 08.30    &-67 50 52.4    &O9.5V                  &14.11  &3\\
D0981201        &Sk-67D250      &MDRS           &05 43 15.48    &-67 51 09.6    &O7.5II(f)              &12.68  &17\\
D0981501        &D301-NW8       &MDRS           &05 43 15.96    &-67 49 51.0    &D301-NW8               &14.37  &3\\
I8120701        &Sk-70D115      &LWRS           &05 48 49.76    &-70 03 57.5    &O6.5Iaf                &12.24  &5\\
\hline
%\end{tabular}
%\end{table}
\end{longtable}
\medskip
\noindent
Notes. Units of right ascension are hours, minutes, and seconds; units of declination are in degrees, arcminutes, and arcseconds.\\
References -- (1) \citet{Walborn77}; (2) \citet{Conti86}; (3) \citet{Massey02}; (4) \citet{Smith96}; (5) \citet{Blair09}; (6) \citet{Walborn95}; (7) \citet{Garmany87} (8) \citet{Rousseau78}; (9) \citet{Moffat90}; (10) \citet{Smith90}; (11) \citet{Walborn02b} (12) \citet{Fitzpatrick88}; (13) \citet{Torres88}; (14) \citet{Walborn02a}; (15) \citet{Walborn97}; (16) \citet{Massey98}; (17) \citet{Massey95} 

\clearpage

%\begin{table}
\begin{longtable}{lccc}
\caption{Equivalent widths and column densities with corresponding velocity limits for O VI absorption in the LMC.}
\label{Tab2}\\

\hline \\[-2ex]
\multicolumn{1}{l}{Target name}  &
\multicolumn{1}{c}{Equivalent width}    &
\multicolumn{1}{c}{log N(O VI)}  &
\multicolumn{1}{c}{Integration limits}\\
\multicolumn{1}{c}{}  &
\multicolumn{1}{c}{(m\AA)}  &
\multicolumn{1}{c}{(dex)}  &
\multicolumn{1}{c}{(km~s$^{-1}$)}\\[0.5ex] \hline \\[-1.8ex]
\endfirsthead

\multicolumn{4}{c}{{\tablename} \thetable{} -- Continued} \\[0.5ex]
\hline \\[-2ex]
\multicolumn{1}{l}{Target name}  &
\multicolumn{1}{c}{Equivalent width}    &
\multicolumn{1}{c}{log N(O  VI)}  &
\multicolumn{1}{c}{Integration limits}\\
\multicolumn{1}{c}{}  &
\multicolumn{1}{c}{(m\AA)}  &
\multicolumn{1}{c}{(dex)}  &
\multicolumn{1}{c}{(km~s$^{-1}$)}\\[0.5ex] \hline \\[-1.8ex]
\endhead

\multicolumn{4}{l}{{Continued on Next Page\ldots}} \\
\endfoot

%\\[-1.8ex] \hline \hline
\endlastfoot
%\caption{Equivalent widths and column densities with corresponding velocity limits for O~{\small VI} absorption in the LMC.}
%\begin{tabular}{l c c c}
%\hline
%Object name     &Equivalent Width       &Log (Col Density)      	&Integration limits\\
%		&(m\AA)			&(dex)				&(km~s$^{-1}$)\\
%\hline
Sk-67D05        &73$\pm$6       	&13.72$^{+0.10}_{-0.13}$        &175, 330\\
Sk-68D03        &112$\pm$8      	&14.02$^{+0.11}_{-0.15}$        &180, 330\\
BI13            &88$\pm$4       	&13.94$^{+0.08}_{-0.09}$        &175, 315\\
Sk-67D18        &132$\pm$7      	&14.16$^{+0.09}_{-0.13}$        &165, 330\\
Sk-67D20        &147$\pm$8      	&14.08$^{+0.11}_{-0.16}$        &175, 335\\
PGMW-3070       &132$\pm$23     	&14.10$^{+0.13}_{-0.18}$        &180, 345\\
LH103102        &132$\pm$23     	&14.16$^{+0.03}_{-0.04}$        &180, 330\\
LH91486         &266$\pm$29     	&14.47$^{+0.09}_{-0.11}$        &175, 385\\
PGMW-3223       &129$\pm$15     	&14.10$^{+0.06}_{-0.07}$        &175, 315\\
HV2241          &271$\pm$18     	&14.50$^{+0.04}_{-0.03}$        &165, 365\\
HD32402         &271$\pm$8      	&14.48$^{+0.02}_{-0.02}$        &160, 320\\
Sk-67D32        &129$\pm$7      	&14.11$^{+0.03}_{-0.04}$        &165, 360\\
Sk-65D21        &94$\pm$23       	&13.94$^{+0.11}_{-0.16}$        &225, 340\\
HV2274          &160$\pm$11     	&14.14$^{+0.10}_{-0.15}$        &165, 345\\
Sk-66D51        &185$\pm$10     	&14.26$^{+0.07}_{-0.09}$        &170, 370\\
NGC1818-D1      &280$\pm$13     	&14.52$^{+0.05}_{-0.05}$        &150, 340\\
Sk-70D69        &182$\pm$12     	&14.18$^{+0.07}_{-0.08}$        &150, 295\\
Sk-67D69        &232$\pm$28     	&14.40$^{+0.05}_{-0.06}$        &170, 340\\
MACHO78-6097    &151$\pm$25      	&14.15$^{+0.05}_{-0.05}$        &165, 320\\
BI130           &94$\pm$7       	&13.89$^{+0.07}_{-0.09}$        &165, 320\\
Sk-69D94        &150$\pm$3      	&14.22$^{+0.03}_{-0.03}$        &160, 340\\
SNR0519-697     &97$\pm$4       	&13.97$^{+0.07}_{-0.08}$        &160, 300\\
BREY22          &126$\pm$24      	&14.07$^{+0.06}_{-0.07}$        &160, 330\\
HD269445        &200$\pm$10     	&14.29$^{+0.04}_{-0.05}$        &175, 365\\
Sk-69D124       &331$\pm$8      	&14.57$^{+0.04}_{-0.05}$        &190, 430\\
Sk-67D105       &110$\pm$8      	&13.92$^{+0.14}_{-0.23}$        &180, 320\\
Sk-67D106       &180$\pm$7      	&14.30$^{+0.10}_{-0.13}$        &175, 345\\
Sk-67D107       &254$\pm$10      	&14.45$^{+0.04}_{-0.05}$        &160, 360\\
HD36521         &126$\pm$9      	&14.07$^{+0.09}_{-0.12}$        &175, 340\\
Sk-68D80        &303$\pm$16     	&14.56$^{+0.04}_{-0.03}$        &145, 335\\
Sk-68D82        &141$\pm$10     	&14.17$^{+0.04}_{-0.04}$        &165, 320\\
BI170           &235$\pm$20      	&14.43$^{+0.02}_{-0.03}$        &165, 365\\
Sk-67D111       &214$\pm$19     	&14.34$^{+0.05}_{-0.05}$        &175, 365\\
HV2543          &156$\pm$42      	&14.23$^{+0.04}_{-0.03}$        &160, 365\\
Sk-70D91        &256$\pm$7      	&14.43$^{+0.04}_{-0.03}$        &160, 365\\
Sk-66D100       &214$\pm$21     	&14.34$^{+0.03}_{-0.03}$        &160, 340\\
HDE269599       &123$\pm$6      	&14.05$^{+0.03}_{-0.03}$        &165, 320\\
Sk-65D63        &184$\pm$8      	&14.21$^{+0.04}_{-0.04}$        &180, 375\\
HV982           &208$\pm$58      	&14.33$^{+0.12}_{-0.17}$        &175, 360\\
Sk-70D97        &226$\pm$4      	&14.39$^{+0.03}_{-0.04}$        &175, 375\\
Sk-67D144       &194$\pm$10     	&14.25$^{+0.06}_{-0.07}$        &165, 335\\
BI184           &118$\pm$10     	&14.04$^{+0.08}_{-0.10}$        &165, 330\\
NGC2004-B15     &92$\pm$7       	&13.86$^{+0.05}_{-0.07}$        &175, 330\\
Sk-71D38        &94$\pm$7       	&13.92$^{+0.07}_{-0.08}$        &165, 315\\
Sk-71D45        &194$\pm$9      	&14.26$^{+0.06}_{-0.06}$        &160, 345\\
Sk-67D166       &206$\pm$9      	&14.32$^{+0.05}_{-0.05}$        &165, 390\\
Sk-67D168       &186$\pm$12     	&14.26$^{+0.05}_{-0.05}$        &165, 375\\
SNR0532-675     &147$\pm$9      	&14.16$^{+0.07}_{-0.08}$        &165, 345\\
Sk-67D191       &229$\pm$16      	&14.42$^{+0.02}_{-0.02}$        &165, 340\\
HV5936          &174$\pm$8      	&14.18$^{+0.07}_{-0.08}$        &175, 375\\
Sk-69D191       &185$\pm$25     	&14.22$^{+0.06}_{-0.07}$        &165, 340\\
J053441-693139  &182$\pm$33      	&14.22$^{+0.05}_{-0.05}$        &165, 330\\
Sk-67D211       &144$\pm$8      	&14.11$^{+0.10}_{-0.13}$        &160, 350\\
BREY64          &139$\pm$8      	&14.20$^{+0.16}_{-0.23}$        &180, 330\\
SNR0536-692     &116$\pm$7      	&14.03$^{+0.05}_{-0.07}$        &165, 320\\
Sk-69D220       &128$\pm$9      	&14.09$^{+0.08}_{-0.11}$        &160, 315\\
Sk-66D172       &173$\pm$9      	&14.20$^{+0.05}_{-0.05}$        &175, 360\\
BI253           &259$\pm$46      	&14.46$^{+0.04}_{-0.05}$        &160, 300\\
Sk-68D137       &234$\pm$7      	&14.45$^{+0.03}_{-0.02}$        &165, 330\\
MK42            &228$\pm$24     	&14.41$^{+0.07}_{-0.08}$        &160, 330\\
Sk-69D243       &307$\pm$15     	&14.56$^{+0.05}_{-0.06}$        &150, 345\\
30DOR-S-R136    &185$\pm$25      	&14.25$^{+0.05}_{-0.06}$        &165, 320\\
Sk-69D246       &211$\pm$7      	&14.37$^{+0.03}_{-0.03}$        &155, 325\\
HDE269927       &245$\pm$8      	&14.42$^{+0.03}_{-0.04}$        &160, 320\\
Sk-69D257       &171$\pm$8      	&14.20$^{+0.07}_{-0.08}$        &160, 315\\
SNR0543-689     &186$\pm$34      	&14.27$^{+0.10}_{-0.12}$        &160, 360\\
D301-1005       &284$\pm$57      	&14.53$^{+0.02}_{-0.03}$        &165, 385\\
Sk-67D250       &316$\pm$33      	&14.57$^{+0.03}_{-0.03}$        &165, 375\\
D301-NW8        &228$\pm$30      	&14.42$^{+0.02}_{-0.03}$        &175, 365\\
Sk-70D115       &186$\pm$11      	&14.23$^{+0.03}_{-0.04}$        &165, 330\\
\hline
%\end{tabular}
\end{longtable}
\medskip
\noindent
Notes. The errors in equivalent widths and column densities are 1$\sigma$ error estimates.\\
Integration limits gives the velocity range over which the LMC O VI absorption profile was integrated.
%\end{table}

\clearpage

\begin{table}
\caption{O~{\small VI} column densities in the superbubbles (SB) of the LMC.}
\label{Tab3}
\begin{tabular}{l c c c}
\hline
Target name    &Superbubble    &Column density 	       &log~N(O~{\small VI})\\
	       &	       &(10$^{14}$~atoms cm$^{-2}$)    &(dex)\\
\hline
Sk-69D246      &30Dor C        &2.36$^{+0.18}_{-0.14}$ &14.37$^{+0.03}_{-0.03}$\\
MK42           &30Dor C        &2.60$^{+0.46}_{-0.45}$ &14.41$^{+0.07}_{-0.08}$\\
30DOR-S-R136   &30Dor C        &1.77$^{+0.20}_{-0.24}$ &14.25$^{+0.05}_{-0.06}$ \\
Sk-69D243      &30Dor C        &3.63$^{+0.40}_{-0.45}$ &14.56$^{+0.05}_{-0.06}$\\
Sk-69D191      &N154           &1.65$^{+0.26}_{-0.23}$ &14.22$^{+0.06}_{-0.07}$\\
HDE269927      &N158           &2.62$^{+0.16}_{-0.21}$ &14.42$^{+0.03}_{-0.04}$\\
Sk-68D80       &N144           &3.61$^{+0.30}_{-0.26}$ &14.56$^{+0.04}_{-0.03}$\\
HD36521        &N144           &1.18$^{+0.28}_{-0.28}$ &14.07$^{+0.09}_{-0.11}$\\
BI184          &N206           &1.09$^{+0.22}_{-0.22}$ &14.04$^{+0.08}_{-0.10}$\\
Sk-71D45       &N206           &1.80$^{+0.24}_{-0.21}$ &14.26$^{+0.06}_{-0.06}$\\
Sk-70D91       &N204           &2.69$^{+0.24}_{-0.17}$ &14.43$^{+0.04}_{-0.03}$\\
PGMW-3223      &N11            &1.27$^{+0.18}_{-0.18}$ &14.10$^{+0.06}_{-0.07}$\\
LH103102       &N11            &1.46$^{+0.10}_{-0.13}$ &14.16$^{+0.03}_{-0.04}$\\
PGMW-3070      &N11            &1.27$^{+0.43}_{-0.43}$ &14.10$^{+0.13}_{-0.18}$\\
LH91486        &N11            &2.96$^{+0.68}_{-0.68}$ &14.47$^{+0.09}_{-0.11}$\\
Sk-67D111      &N51            &2.20$^{+0.29}_{-0.24}$ &14.34$^{+0.05}_{-0.05}$\\
Sk-67D107      &N51            &2.85$^{+0.27}_{-0.27}$ &14.45$^{+0.04}_{-0.05}$\\
Sk-67D106      &N51            &2.01$^{+0.50}_{-0.53}$ &14.30$^{+0.10}_{-0.13}$\\
Sk-67D166      &N57            &2.09$^{+0.25}_{-0.20}$ &14.32$^{+0.05}_{-0.05}$\\
Sk-67D250      &N70            &3.71$^{+0.23}_{-0.23}$ &14.57$^{+0.03}_{-0.03}$\\
D301-1005      &N70            &3.37$^{+0.17}_{-0.22}$ &14.53$^{+0.02}_{-0.03}$\\
D301-NW8       &N70            &2.60$^{+0.09}_{-0.13}$ &14.42$^{+0.02}_{-0.03}$\\
Total SB (this work) &         &2.31$\pm$0.12	       &14.35$\pm$0.18\\
Total non-SB (this work) &     &1.68$\pm$0.12	       &14.19$\pm$0.23\\
Total SB N70$^{1}$ &           &3.04$\pm$0.45  	       &14.48$\pm$0.06\\
Total non-SB$^{1}$ &           &2.03$\pm$0.58          &14.29$\pm$0.14\\
Total SB$^{2}$   &             &3.07$\pm$0.62          &14.48$\pm$0.09\\
Total non-SB$^{2}$ &           &2.10$\pm$0.50          &14.31$\pm$0.11\\
\hline
\end{tabular}
\end{table}
\medskip
\noindent
Notes. The superbubble lines of sight are shown in Fig. \ref{Fig1}.\\
(1) \citet{Danforth06a}; (2) Combined data of \citet{Howk02a} and \citet{Danforth06a}.

\end{document}